\documentstyle[amsfonts,epsfig,12pt,cite]{article}
\newcommand{\vb}[1]{{\mathbf{#1}}}
\newcommand{\r}[1]{\ref{#1}}
\newcommand{\lb}[1]{\label{#1}}

\newcommand{\bc}{\begin{center}}
\newcommand{\ec}{\end{center}}
\newcommand{\be}{\begin{equation}}
\newcommand{\ee}{\end{equation}}
\newcommand{\bea}{\begin{eqnarray}}
\newcommand{\eea}{\end{eqnarray}}
\newcommand{\ba}[1]{\begin{array}{#1}}
\newcommand{\ea}{\end{array}}
\newcommand{\bz}{{\overline{z}}}

\newcommand{\bt}[1]{\begin{table}[ht]\centering\begin{tabular}{#1}}
\newcommand{\et}[1]{\end{tabular}\caption{\small#1}\end{table}}
\newcommand{\fig}[3]{\begin{figure}[htb]\epsfxsize=100mm\bigskip
\centerline{\epsfbox{#1}}\caption{\small\it #2 \label{#3}}\bigskip\end{figure}}
\newcommand{\figbig}[3]{\begin{figure}[htb]\epsfxsize=100mm\bigskip
\centerline{\epsfbox{#1}}\caption{\small\it #2 \label{#3}}\bigskip\end{figure}}
\newcommand{\mod}{\,{\mathrm{mod}}\,}

\newcommand{\id}{{1\!\!1}} 
\def\e{{\,\rm e}\,}

\begin{document}


\thispagestyle{empty}

\begin{flushright}

{\small SUSX--TH/01--040 \ \ \
OUTP--01--57P \ \ \
IHES/P/01/53\\
LPT--Orsay--01/114 \ \ \
HWM--01--40 \ \ \
EMPG--01--19\\
{\tt hep-th/0112104} \ \ \ {\sl December 2001}}

\end{flushright}

\begin{center}

{\Large\bf{Conformal Orbifold Partition Functions\\[2mm] from Topologically
Massive Gauge Theory}}\\

\vspace{1 truecm}

{\bf P. Castelo Ferreira}
\\[2mm]
{\small{\it Center for Theoretical Physics -- University of Sussex,
Falmer, Brighton BN1 9QJ, U.K.} and\\ {\it Dep. de Matem\'atica -- Instituto
Superior T\'ecnico, Av. Rovisco Pais, 1049-001 Lisboa, Portugal}}\\ {\tt
pcf20@pact.cpes.susx.ac.uk}
\\[5mm]
{\bf I.I.\ Kogan}
\\[2mm]
{\small{\it Dept. of Physics, Theoretical Physics -- University of Oxford,
Oxford OX1 3NP, U.K.,\\ IHES, 35 Route de Chartres, 91440 Bures-sur-Yvette,
France,} and\\
{\it Lab. de Physique Th\'eorique, Universit\'e de Paris XI, 91405 Orsay
C\'edex, France }}\\{\tt kogan@thphys.ox.ac.uk}
\\[5mm]
{\bf R.J. Szabo}
\\[2mm]
{\small\it Dept. of Mathematics -- Heriot-Watt University, Riccarton, Edinburgh
EH14 4AS, U.K.}\\{\tt R.J.Szabo@ma.hw.ac.uk}
\\[10mm]

{\bf\sc Abstract}

\begin{minipage}{15cm}
\vspace{2mm}

We continue the development of the topological membrane approach to open and
unoriented string theories. We study orbifolds of topologically massive gauge
theory defined on the geometry $[0,1]\times\Sigma$, where $\Sigma$ is a generic
compact Riemann surface. The orbifold operations are constructed by gauging the
discrete symmetries of the bulk three-dimensional field theory. Multi-loop
bosonic string vacuum amplitudes are thereby computed as bulk correlation
functions of the gauge theory. It is shown that the three-dimensional
correlators naturally reproduce twisted and untwisted sectors in the case of
closed worldsheet orbifolds, and Neumann and Dirichlet boundary conditions in
the case of open ones. The bulk wavefunctions are used to explicitly construct
the characters of the underlying extended Kac-Moody group for arbitrary genus.
The correlators for both the original theory and its orbifolds give the
expected modular invariant statistical sums over the characters.

\end{minipage}

\end{center}

\noindent

\vfill
\begin{flushleft}
PACS: 11.10.Kk, 11.15.Tk, 11.25.Hf, 11.25.Sq\\
Keywords: Chern-Simons theories, Strings, Conformal field theory, Orbifolds,
Statistical sums\\
\end{flushleft}

\newpage

\addtolength{\baselineskip}{0.20\baselineskip}

\thispagestyle{empty}
\tableofcontents

\newpage

\setcounter{page}{1}

\setcounter{equation}{0}

\section{Introduction}

This paper is a continuation of the ongoing development of the topological
membrane approach to open and unoriented string theories, in which the focus
will be on obtaining the appropriate spectrum and amplitudes of worldsheet
orbifold models directly from a three-dimensional topologically massive gauge
theory. As we will see, this formalism provides a systematic and powerful way
of constructing correlation functions in two-dimensional conformal field
theories. At the same time, we will obtain a very simple description of
conformally invariant twist sectors and boundary conditions. We thereby
formulate the intrinsic properties of open and unoriented string theories
within the much simpler setting of three-dimensional quantum gauge theory.

\subsection{Prelude}

Although open strings were originally considered as defining theories by
themselves, it soon become evident that they must come accompanied by closed
(non-chiral) strings, in order to maintain unitarity of the string theory.
Furthermore, open string theories may be obtained from closed ones by gauging
certain discrete geometrical symmetries of the closed string
theory~\cite{SBH_00,SBH_00a,SBH_00b,SBH_00c,SBH_00d,SBH_01,SBH_02,SBH_03,SBH_04,SBH_05}. The simplest instance uses worldsheet parity $\Omega:z\mapsto-\bz$ which imposes the identification $\sigma_1\equiv-\sigma_1$ of the worldsheet coordinates, where $z=\sigma_1+i\sigma_2$ and $\bz=\sigma_1-i\sigma_2$ define the complex structure of the worldsheet. The spaces obtained in this way can be either closed and unoriented, or open and oriented (and unoriented as well). The latter worldsheets are generally called orbifolds and the singular points of the orbifold projection become the boundaries of the open string worldsheet. The physical states (or more generally the quantum fields) of the open or unoriented theory are obtained from those of the closed oriented theory by projecting out the ones which have negative worldsheet parity eigenvalues. This is accomplished by building a suitable projection operator that retains only the states of even parity in the theory.

Another construction in string theory involves orbifolding the target space of
the theory with respect to an involution of some spacetime symmetry. In this
paper we will only consider involutions which are of order 2, obtained by
imposing the identifications $X^{I}\equiv-X^{I}$ on the target space
coordinates. Upon combining both worldsheet and target space orbifold
constructions, i.e. $X^{I}(z,\bz)\equiv-X^{I}(-\bz,-z)$, we obtain open or
unoriented theories in orbifolds~\cite{SBH_03,SBH_04,SBH_05}. In other words,
one arrives at orientifolds, leading to twisted sectors in the Hilbert space of
the open or unoriented theories. An important result implied by these
constructions is that the gauge group of the open string theory, describing the
Chan-Paton degrees of freedom carried by the target space photon Wilson lines,
is constrained~\cite{SBH_06,SBH_07,SBH_07a,SBH_07b,SBH_07c}. This constraint is
closely related to the types of boundary conditions that one must impose on
insertions of boundary vertex operators at the level of the worldsheet field
theory~\cite{POL_1}.

The perturbation expansion of a string theory is a sum over the genus $g$ of
the worldsheet $\Sigma$. Each term in the series yields string scattering
amplitudes for a particular order $g$. The object one needs to compute to deal
with the genus expansion is the partition function, or in other words the
vacuum-to-vacuum amplitudes order by order, i.e. for each genus $g$. In this
paper we will compute these objects using the topological membrane approach to
string
theory~\cite{TM_00,TM_01,TM_02,TM_03,TM_04,TM_04a,TM_05,TM_06,TM_07,TM_08,TM_10,TM_11,TM_12,TM_13,TM_14,TM_15,TM_16,TM_17,TM_last}, which consists of
replacing the string worldsheet by a three-dimensional manifold. Dynamically,
it is modelled by a topologically massive gauge
theory~\cite{TMGT_01,TMGT_02,TMGT_03,TMGT_04}, i.e. the gauge theory whose
action contains both a Yang-Mills and a Chern-Simons term for the gauge
fields living on a three-manifold, together with topologically massive
gravity, i.e. the three-dimensional gravitational theory whose action
contains both an Einstein and a gravitational Chern-Simons term. The membrane
is the three-dimensional manifold $M=[0,1]\times\Sigma$ which has two
disjoint boundaries $\partial M=\Sigma_0\amalg\Sigma_1$. The gauge
non-invariance of the resulting theory induces chiral conformal field
theories on the boundaries. The advantage of this approach is that many
stringy properties have very transparent and simple physical
characterizations in terms of three-dimensional gauge and gravity theories.
The original work on this subject concerned only pure Chern-Simons theories
in the bulk~\cite{W_0,TMGT_03,TMGT_04,BN_1,LR_1,O_1,W_1,H_1}.

In this setting, closed string theories are obtained as the effective boundary
theories with worldsheet the closed surface $\partial M$. Obtaining an open
string theory raises a problem, because in that case one needs an open
worldsheet. For a smooth manifold the boundary of a boundary is empty,
$\partial \partial M=\emptyset$, and so naively it seems that topological
membrane theory cannot describe open strings since the induced worldsheets are
already the boundary of a three-manifold. The solution to this problem is to
consider an orbifold of the bulk theory. The fixed points of the orbifold then
play the role of the one-dimensional boundary of the two-dimensional boundary
of the three-dimensional membrane. This approach was first suggested
in~\cite{H_1} within the context of pure Chern-Simons gauge theories, and
developed further in~\cite{TM_16,TM_17}. There the discrete symmetries of the
three-dimensional theory, namely $\Omega T$ and $\Omega CT$, are used as the
generators of the orbifold group.

Various other works have also gone in this direction. Recent studies on
Wess-Zumino-Novikov-Witten (WZNW) orbifold constructions can be found
in~\cite{WZW_1,WZW_2,WZW_3}. In~\cite{branes_0}, D-branes are described at the
worldsheet level as vertex operators. Studies of generic rational conformal
field theories with boundaries from pure three-dimensional Chern-Simons theory
appear in~\cite{FS_1,FS_2,FS_3}. Recent developments related to the topics
covered in this paper may also be found
in~\cite{PS_1,PS_2,PS_3,branes_1,branes_2}.

\subsection{Summary of Results}

This paper continues and develops further the study started
in~\cite{TM_16,TM_17}. The main objective is to compute the partition functions
of two-dimensional open and unoriented conformal field theories from the
perspective of a three-dimensional topological membrane. For simplicity, we
will consider here only a $U(1)$ topologically massive gauge theory whereby
everything can be worked out explicitly. This corresponds to the simplest
instance of bosonic strings compactified on a circle, although the techniques
introduced and conclusions reached also apply to more complicated situations.
In particular, we will reproduce the standard modular invariants corresponding
to a $U(1)$ affine Lie algebra. Some of these results were announced
in~\cite{TM_17} where a general review may also be found.

In the following we will interpret the boundary partition function as the bulk
vacuum-to-vacuum amplitude. This translates into computing boundary-to-boundary
correlators $\left<\Psi_0|\Psi_1\right>$ of ground state
wavefunctions~\cite{BN_1,LR_1}. The wavefunctions of topologically massive
gauge theory~\cite{AFC_1}, inserted at each of the boundaries, work as chiral
WZNW models~\cite{W_1} and render the theory well-defined by effectively
enforcing the appropriate boundary conditions on the full three-dimensional
gauge theory~\cite{TMGT_03,TM_17}. One also has to consider the monopole
effects described in~\cite{TM_15} in order to get the correct
holomorphic-antiholomorphic pairing of the chiral conformal blocks. However,
these boundary insertions must be suitably modified, because we want to
orbifold the bulk theory, but the wavefunctions only live at the boundaries.
The solution is to extend them appropriately to the full three-dimensional
spacetime. Once this is accomplished we can generate the orbifold with a new
boundary at the orbifold plane, and thereby obtain a new wavefunction which we
show has to be the constant state $|1\rangle$ such that the correlation
function becomes $\left<\Psi_0|1\right>$.

In the case that the orbifold boundary is an open surface, the new induced
wavefunction is not quite unity. There are also induced factors resulting from
the functional integration on the new surface which are responsible for setting
the correct conformal boundary conditions for each orbifold, namely Dirichlet
for $\Omega T$ and Neumann for $\Omega CT$. It is necessary in this description
to consider a one-dimensional boundary correction to the Gauss law upon
orbifolding. It is also through this term that insertions of vertex operators
in the induced conformal field theory~\cite{branes_0,branes_1,branes_2} can be
described by the topological membrane~\cite{progress}, particularly for
Dirichlet boundary conditions since in that case the source terms at the
orbifold fixed points couple to the gauge field and act as vertex operators on
the boundary. In fact, this constitutes the first step towards the
construction of D-branes from topological membrane theory~\cite{progress}.
In this way we will explicitly compute, for any genus $g$, the
string vacuum amplitudes entirely from the topological membrane. They are
expressed as modular invariant combinations of affine characters of the
two-dimensional conformal field theory, which we compute from the
three-dimensional gauge theory by appropriately incorporating the remaining
homology dependence of the wavefunctions on the Riemann surface $\Sigma$. An
unsolved problem that remains concerns the construction of more general modular
invariants. Within the present framework this is is not possible without a
mechanism which creates a tunnelling effect between various ground state
wavefunctions that is more general than the one described in~\cite{TM_15}. For
this, it is most likely necessary to take into account the effects of gravity
in the bulk theory~\cite{TM_04a,W_2,TM_last}.

The organization of the remainder of this paper is as follows. In the next
section we review the essential features of topologically massive gauge
theories that we shall need and outline how they will be used to construct the
modular invariant partition functions of conformal field theories. In
section~\ref{sec.orbi} we summarize the three-dimensional orbifold operations
that will be used and how they will naturally induce the amplitudes appropriate
to the orbifold conformal field theories. In section~\ref{sec.wave} we
explicitly construct the vacuum wavefunctions of topologically massive gauge
theory and use them to study the bulk partition function. In
section~\ref{sec.path} we then consider the effects of orbifolding on these
wavefunctions and amplitudes. Finally, in section~\ref{sec.character} we
carefully examine the topological dependence of the gauge theory wavefunctions
and derive very explicit expressions for the conformal field theory modular
invariants valid for arbitrary genus $g$. The case of genus one is studied in
some detail for illustration.

\setcounter{equation}{0}

\section{\lb{sec.tmgt}Canonical Formalism for Topologically Massive Gauge
Theories}

In this section we will review some of the basic aspects of $U(1)$
topologically massive gauge theories~\cite{TMGT_01,TMGT_02,TMGT_03,TMGT_04}
that will be required in the following.
The gauge theory is defined on a three-dimensional manifold of the form
$M=[0,1]\times\Sigma$, where the finite interval $[0,1]$ parametrizes the time
$t$, and $\Sigma$ is a compact Riemann surface whose local complex coordinates
will be denoted $\vb{z}=(z,\bz)$ with integration measure $d^2z=|dz\wedge
d\bz|$. We will adopt Gaussian normal coordinates for the Minkowski
three-geometry, in which the metric takes the form
\be
ds_{(3)}^2=g_{\mu\nu}~dx^\mu~dx^\nu=-dt^2+h_{ij}~dx^i~dx^j \ .
\label{ds3}\ee
The three-manifold has two boundaries $\Sigma_0$ and $\Sigma_1$ at times $t=0$
and $t=1$, respectively, both of which are copies of $\Sigma$ with opposite
orientation. The action is a sum of Maxwell and Chern-Simons terms for a
$U(1)$ gauge field $A$ with curvature $F$,
\be
S_{\rm TMGT}[A]=\int\limits_0^1dt~\int\limits_\Sigma d^2z~
\left[-\frac{\sqrt{-g}}{4\gamma}\,F_{\mu\nu}F^{\mu\nu}+\frac{k}{8\pi}\,
\epsilon^{\mu\nu\lambda}\,A_\mu\,\partial_\nu A_\lambda+
\sqrt{-g}\,A_\mu J^\mu\right] \ ,
\lb{STMGT}
\ee
where we have included the minimal coupling of the gauge fields to a conserved
current $J^\mu$,
\be
\partial_\mu J^\mu=0 \ .
\label{Jmuconserved}
\ee
The bulk Levi-Civita antisymmetric tensor density $\epsilon^{\mu\nu\lambda}$
induces the tensor density $\epsilon^{ij}=\epsilon^{0ij}$ on the boundaries. As
has been extensively studied in the
past~\cite{TM_00,TM_01,TM_02,TM_03,TM_04,TM_05,TM_06,TM_07,TM_08,TM_10,TM_11,TM_12,TM_13,TM_14,TM_15,TM_16,TM_17,TM_last,W_0,BN_1,LR_1,O_1}, the quantum
field theory defined by (\ref{STMGT}) induces new degrees of freedom on the
boundaries which constitute chiral WZNW models. They are fields belonging to
two-dimensional chiral conformal field theories living on $\Sigma_0$ and
$\Sigma_1$.

\subsection{\lb{subsec.canquant}Hamiltonian Quantization}

The action (\ref{STMGT}) can be written in a canonical splitting
$A_\mu=(A_0,A_i)$ as
\be
\ba{lll}
S_{\rm TMGT}[A]&=&\displaystyle\int\limits_0^1dt~\int\limits_\Sigma d^2z~
\left[-\frac{\sqrt{-g}}{2\gamma}\,F_{0i}F^{0i}-\frac{\sqrt{-g}}{4\gamma}\,
F_{ij}F^{ij}+\frac{k}{16\pi}\,\epsilon^{ij}\,A_0F_{ij}+\frac{k}{8\pi}\,
\epsilon^{ij}\,A_iF_{j0}\right.\\[2mm]&&\displaystyle+\biggl.\sqrt{-g}\,A_0J^0+
\sqrt{-g}\,A_iJ^i\biggr] \ .
\ea
\ee
The canonical momentum conjugate to $A_i$ is
\be
\Pi^i=-\frac{\sqrt{-g}}{\gamma}\,F^{0i}+\frac{k}{8\pi}\,\epsilon^{ij}\,A_j \ ,
\ee
while, as usual, the canonical momentum conjugate to $A_0$ is identically
zero. The field $A_0$ is therefore non-dynamical and serves as a Lagrange
multiplier which imposes the Gauss law constraint
\be
0=\int\limits_{\Sigma}
d^2z~\left(-\frac{\sqrt{h}}{\gamma}\,\partial_iF^{0i}+\frac{k}{8\pi}\,
\epsilon^{ij}\,F_{ij}+\sqrt h\,J^0\right)-\oint\limits_{\partial\Sigma}
\left(-\frac{\sqrt{h}}{\gamma}\,F^{0i}+\frac{k}{8\pi}\,\epsilon^{ij}\,
A_{j}\right)n_i \ ,
\lb{gauss}
\ee
where $n_i$ is a vector normal to the boundary of $\Sigma$. Note that the
boundary term in (\ref{gauss}) is only present when the two-dimensional
boundary $\Sigma$ of the underlying three-manifold has a boundary. Of course
for a smooth space this term doesn't appear because the boundary of a boundary
is empty. However, once we quotient the theory by its discrete symmetries new
boundaries can emerge at orbifold singularities~\cite{H_1,TM_16,TM_17}. This
extra boundary term is vital for the construction we will present in the
following, because it allows the imposition of the correct boundary conditions
on the conformal field theory. Moreover, conformal vertex operators inserted on
the boundary are thereby included in the full three-dimensional theory as
\textit{external} fluxes coupled to the gauge fields through the conserved
current $J^\mu$~\cite{progress}, in accordance with the fact that closed string
vertex operators correspond to Wilson lines of the three-dimensional gauge
theory. We will see later on that the external charges actually allow one to
introduce collective coordinates of D-branes~\cite{progress}.
It is also this mechanism that constrains the open string gauge group
and therefore the Chan-Paton degrees of freedom~\cite{SBH_07,SBH_07a,SBH_07b}.

The Hamiltonian of the theory is given by
\be
\ba{rcl}
H&=&\displaystyle\int\limits_\Sigma d^2z~\left\{
-A_0\left[\partial_i\left(\Pi^i-\frac{k}{8\pi}\,
\epsilon^{ij}\,A_j\right)+\frac{k}{4\pi}\,\epsilon^{ij}\,\partial_iA_j
+\sqrt h\,J^0\right]\right.\\[2mm]&&\displaystyle+\,
\partial_i\left[A_0\left(\Pi^i+\frac k{8\pi}\,\epsilon^{ij}\,A_j\right)
\right]+\frac{1}{8\gamma\,\sqrt{h}}\,\left(\epsilon^{ij}\,
F_{ij}\right)^2\\[2mm]&&\displaystyle+\left.
\frac{\gamma}{2\,\sqrt{h}}\,h_{ij}\left(\Pi^i-\frac{k}{8\pi}\,
\epsilon^{ik}\,A_k\right)\left(\Pi^j-\frac{k}{8\pi}\,\epsilon^{jl}\,A_l\right)
-\sqrt h\,A_iJ^i\right\} \ .
\label{Ham}
\ea
\ee
Defining the electric and magnetic fields as
\be
\ba{rcl}
E^i&=&-\displaystyle\frac{1}{\gamma}\,F^{0i} \ , \\[4mm]
B  &=&\partial_z A_\bz-\partial_\bz A_z \ ,
\ea
\ee
the Gauss law constraint (without boundary terms) reads
\be
\partial_iE^i+\frac{k}{4\pi}\,B=-J^0 \ .
\label{GaussEB}\ee
In the quantum field theory, the canonical commutation relations can be written
as
\be
\ba{rcl}
\Bigl[E^i(\vb{z})\,,\,E^j(\vb{z}')\Bigr]&=&\displaystyle-i\,{k\over{4\pi}}
\,\epsilon^{ij}\,\delta^{(2)}(\vb{z}-\vb{z}') \ , \\[2mm]
\Bigl[E^i(\vb{z})\,,\,B(\vb{z'})\Bigr]&=&-i\,\epsilon^{ij}\,\partial_j
\delta^{(2)}(\vb{z}-\vb{z}') \ ,
\ea
\lb{com_EB}
\ee
and the constraint (\ref{GaussEB}) leads to an equation that needs to be
satisfied by the physical (gauge invariant) states. We will use this equation
in section~\ref{sec.wave} when we construct the quantum wavefunctions of the
field theory. Note that the $A_0$ dependent terms in (\ref{Ham}) vanish when
$H$ acts on such states.

The generators of time-independent local gauge transformations can be easily
defined, for smooth real-valued gauge parameter functions $\Lambda$,
as~\cite{TM_07}
\be
U_\Lambda=\exp\left\{i \int\limits_\Sigma d^2z~\sqrt h\,\Lambda(\vb{z})
\left(\partial_iE^i+{k\over{4\pi}}\,B+J^0\right)\right\} \ .
\lb{U}
\ee
The physical Hilbert space consists of those quantum states which are invariant
under the actions of the operators (\ref{U}). In addition, when there are
topologically non-trivial gauge field configurations, we must take into account
the large gauge transformations of the theory. These are generated by the
operators (\ref{U}) obtained by taking $\Lambda$ to be the multi-valued angle
function of the Riemann surface $\Sigma$. Then integrating by parts in
(\ref{U}) yields the extra local operator~\cite{V_01,V_02,V_03,V_04}
\be
V(\vb{z}_0)=\exp\left\{-i\int\limits_\Sigma d^2z~\left[\left(E^i+
\frac{k\,\sqrt h}{4\pi}\,\epsilon^{ij}\,A_j\right)\epsilon_{ik}\,\partial^k
\ln E(\vb{z},\vb{z}_0)-\sqrt h\,
\theta(\vb{z},\vb{z}_0)\,J^0\right]\right\} \ ,
\lb{V}
\ee
where $E(\vb{z},\vb{z}_0)$ is the prime form of $\Sigma$ (see
section~\ref{subsec.CFTpart}), $\vb{z}_0$ is a fixed point on $\Sigma$, and
\be
\theta(\vb{z},\vb{z}_0)={\rm Im}\,\ln\frac{E(\vb{z},\vb{z}_0)}
{E(\vb{z},\vb{z}')E(\vb{z}',\vb{z}_0)}
\label{thetaImdef}\ee
with $\vb{z}'$ an arbitrary fixed reference point. Demanding invariance under
these operators, i.e. under large gauge transformations, further truncates the
physical Hilbert space of the quantum field theory. Moreover, from the
commutation relations~(\r{com_EB}) we obtain the relation
\be
\Bigl[B(\vb{z})\,,\,V^n(\vb{z}_0)\Bigr]=2\pi n\,\sqrt h\,V^n(\vb{z}_0)\,
\delta^{(2)}(\vb{z}-\vb{z}_0)
\ee
for any integer $n$. This means that the operator $V^n$ creates a pointlike
magnetic vortex at $\vb{z}_0$ with magnetic flux $\int_\Sigma d^2z~\sqrt h\,
B=2\pi n$. These objects thereby generate nonperturbative processes which
constitute monopoles of the gauge theory. Moreover, from Gauss' law
(\ref{GaussEB}) we see that they also carry a bulk electric charge
\be
\Delta Q=-\frac{nk}2 \ .
\ee

The electric charge spectrum of the quantum field theory is~\cite{TM_07}
\be
Q_{m,n}=m+\frac{k}{4}\,n \ ,
\lb{charge}
\ee
where $m$ and $n$ are integers representing, respectively, the contributions
from the usual Dirac charge quantization and the monopole flux. Due to the
existence of monopole induced processes and linkings between Wilson lines
(charge trajectories) it can be shown~\cite{TM_15} that, with the correct
relative boundary conditions, the insertion of the charge $Q_{m,n}$ at one
boundary $\Sigma_0$ (corresponding to a vertex operator insertion in the
boundary conformal field theory) necessitates an insertion of the charge
\be
\bar{Q}_{m,n}=m-\frac{k}{4}\,n
\lb{ccharge}
\ee
at the other boundary $\Sigma_1$. This fact will be assumed throughout the rest
of this paper.

\subsection{\lb{subsec.bdrypart}Boundary Partition Functions}

In the following we will explain how to explicitly build orbifolded boundary
partition functions of conformal field theory from correlation functions in the
three-dimensional gauge theory. For this, we will rederive the well-known
one-to-one correspondence between the wavefunctions of the gauge theory
and the conformal blocks of the boundary conformal field
theory~\cite{W_0,BN_1,LR_1}. In particular, for rational values
\be
k=\frac{2p}q
\ee
of the Chern-Simons coefficient, with $p$ even, there are $(pq)^g$ independent
wavefunctions corresponding to the conformal blocks of the induced $c=1$
rational conformal field theory, i.e. of the WZNW model that emerges on the
boundary $\Sigma$, where $g$ is the genus of $\Sigma$. The gauge theory
partition function then takes the form
\be
Z_{\rm TMGT}=\sum_{\lambda\in({\mathbb{Z}}_{pq})^g}\,
\Bigl\langle\Psi_\lambda(0)\Bigm|\Psi_{\lambda}(1)\Bigr\rangle \ ,
\lb{Z}
\ee
where $\Psi_\lambda(0)$ and $\Psi_\lambda(1)$ denote the wavefunctions at the
boundaries $\Sigma_0$ and $\Sigma_1$, respectively. It is the correct
definition of the object (\ref{Z}) and its orbifold version that we are going
to deal with in this paper.

As explained in the previous subsection, a vertex operator insertion
corresponding to a collection of primary charges
\be
\frac{\lambda^\ell}q=m^\ell+\frac{kn^\ell}4 \ , ~~ \ell=1,\dots,g
\label{lambdadef}
\ee
at one boundary will correspond at the other boundary to an insertion with
charges
\be
\frac{\bar{\lambda}^\ell}q=m^\ell-\frac{kn^\ell}4 \ , ~~ \ell=1,\dots,g \ ,
\label{barlambdadef}
\ee
where $m^\ell$ and $n^\ell$ are winding and monopole numbers acquired by a
charge as it moves around the canonical homology cycles of the Riemann surface
$\Sigma$. Here and in the following we will exploit the fact that a
monopole singularity effectively produces a puncture on $\Sigma$ and hence
extra non-contractible loops. For ease of notation, we will, without loss of
generality, model the extra homology generators using a set of canonical cycles
of the surface. The charge non-conserving processes are then determined by the
large gauge transformations which wind around these cycles. In terms of the
wavefunctions (or conformal blocks) this raises a problem since we have not yet
determined how to introduce the monopole processes into the path integral. They
can be thought of as insertions of the local Gauss operator $V$ given
in~(\r{V}). The main point is that the physical wavefunctions are the ones
which are invariant under the actions of all combinations of the operators
$V$~\cite{V_01,V_02,V_03,V_04}.\footnote{Thanks to Alex Kovner
for this useful remark.}
Generally, physical states are constructed from a generic one $|\Psi\rangle$
through a projection onto gauge invariant states, which may be realized
explicitly by integrating over all gauge copies of the state,
\be
{\cal P}|\Psi\rangle=\frac1{{\rm vol}\,{\cal G}}\,\int D\Lambda
{}~U_\Lambda|\Psi\rangle \ ,
\label{gaugeproj}
\ee
where $\cal G$ is the group of gauge transformations $\Sigma\to U(1)$. This
means, in particular, that one must include arbitrary monopole processes in all
physical amplitudes. Moreover, as shown in~\cite{TM_15}, the total probability
amplitude for the combinations of wavefunctions (at the boundaries) which are
not invariant under these operators vanishes.

In this way the effective path integral will have to take into account this
phenomenon and be of the form
\be
\ba{lll}
\displaystyle Z_\Psi&=&\displaystyle
\sum_{\lambda\in({\mathbb{Z}}_{pq})^g}\,\Bigl\langle\Psi_\lambda(0)
\Bigm|\prod_{a} V^{n_a}(\vb{z}_a)\,\sum_{\lambda'\in({\mathbb{Z}}_{pq})^g}
\Bigm|\Psi_{\lambda'}(1)\Bigr\rangle\\[4mm]
&=&\displaystyle
\sum_{\lambda\in({\mathbb{Z}}_{pq})^g}\,\Bigl\langle\Psi_\lambda(0)
\Bigm|\Psi_{\bar{\lambda}}(1)\Bigr\rangle \ .
\ea
\lb{ZV}
\ee
We will not develop this issue in detail here, as it is enough to consider
shifts of the charges at $\Sigma_1$ by amounts $\Delta Q^\ell=-kn^\ell/2$ or
$\Delta Q^\ell=0$, depending on the relative boundary conditions, with respect
to those at $\Sigma_0$. The effective wavefunction at the boundary $\Sigma_1$
is then given in terms of a Wilson loop as
\be
\Psi_{\bar{\lambda}}(1)=\exp\left\{i\,\sum_{\ell=1}^g\Delta Q^\ell\,
\oint\limits_{\beta^\ell}A\right\}\,\Psi_{\lambda}(0)
\lb{Psi-W}
\ee
with $\lambda^\ell/q=m^\ell+n^\ell k/4$ and
$\bar\lambda^\ell/q=\lambda^\ell/q+\Delta Q^\ell$, where the contour integrals
go around canonical homology cycles $\beta^\ell$ associated with the vortex
sources on $\Sigma$. These nonperturbative processes, i.e. braidings of Wilson
lines and monopole transitions, will take into account the difference in
electric charges between the two boundaries $\Sigma_0$ and $\Sigma_1$.

The free boundary partition function is then of the schematic form
\be
Z_{\mathrm{free}}=\sum_{\lambda\in({\mathbb{Z}}_{pq})^g}\,
\Bigl\langle\Psi_\lambda(0)\Bigm|
\Psi_\lambda(1)\Bigr\rangle+\sum_{\lambda\in({\mathbb{Z}}_{pq})^g}\,
\Bigl\langle\Psi_\lambda(0)\Bigm|\Psi_{\bar\lambda}(1)\Bigr\rangle \ .
\ee
This is the simplest modular invariant of the boundary conformal field theory.
Depending on the relative boundary conditions of the three-dimensional theory,
we can set it to be either only the first sum or only the second sum. For free
boundary conditions, both sums are present. Generic modular invariants would
arise as bilinear combinations
\be
Z_\zeta=\sum_{\lambda,\lambda'}\zeta^{~}_{\lambda\lambda'}\,\Bigl\langle
\Psi_\lambda(0)\Bigm|\Psi_{\lambda'}(1)\Bigr\rangle
\ee
of the wavefunctions with non-negative integral coefficients
$\zeta^{~}_{\lambda\lambda'}$. The allowed combinations are specified by the
requirement of a modular invariant partition function. Such a classification
generally requires the inclusion of the gravitational sector of the topological
membrane, i.e. the coupling of topologically massive gauge theory to
topologically massive gravity. This analysis would take us beyond the scope of
the present paper, and thus we shall focus only on the basic modular
invariants~(\ref{ZV}).

\setcounter{equation}{0}

\section{\lb{sec.orbi}Orbifold Constructions}

In~\cite{TM_16} the procedure to orbifold the topological membrane by
gauging the discrete symmetries of topologically massive gauge theory was
introduced. In this setting, the original worldsheet $\Sigma$ is regarded as a
double cover of an open or unoriented surface $\Sigma^O$, which is obtained
formally as the quotient $\Sigma^O=\Sigma/{\mathbb{Z}}_2$ by an anti-conformal
involution $O$ whose fixed points correspond to the boundary points of
$\Sigma^O$. Of course if $\Sigma^O$ is itself oriented and closed, the cover
consists of the two disjoint copies $\Sigma_0$ and $\Sigma_1$ of $\Sigma$
endowed with opposite orientation. When the worldsheet involution is combined
with a time reversal operation, we may then regard the worldsheet $\Sigma^O$
from the topological membrane perspective as the connecting three-manifold
$\Bigl([0,1]\times\Sigma\Bigr)/{\mathbb{Z}}_2$. If $\Sigma^O$ is closed and
orientable, then the connecting three-manifold is a cylinder over $\Sigma^O$.

Let us recall the discrete symmetries of the theory defined on the geometry
$[0,1]\times\Sigma$. They are time reversal $T: t\mapsto
1-t,\mathbf{A}\mapsto\mathbf{-A}$, charge conjugation $C:A_\mu\mapsto -A_\mu$,
and worldsheet parity $\Omega:z\mapsto-\bz$. It is also possible to introduce
other definitions of parity, which would lead to different types of orbifold
constructions. Due to the presence of both the Maxwell and Chern-Simons terms
in the action (\ref{STMGT}), the only possible combinations of these three
symmetries which are compatible with the three-dimensional action are $\Omega
CT$ and $\Omega T$. Note that only the worldsheet $\Sigma$ feels the parity
operation. For example, consider the case of a torus $\Sigma=T^2$ of modulus
$\tau=\tau_1+i\tau_2$. Orbifolding it with $\tau=i\tau_2$ under $\Omega$
produces an annulus (or a cylinder), while for $\tau=1/2+i\tau_2$ a M\"obius
strip is obtained. On the other hand, starting with $\tau=2i\tau_2$ and the
orbifold under $\Omega':z\mapsto-\bz+2\pi i \tau_2$, a Klein bottle is
produced. See~\cite{TM_16} for further details. We will return to these
examples in section~\ref{sec.character}.

We will use the following notation in the remainder of this paper. We assume a
decomposition of the fields into components parallel ($\parallel$) and
perpendicular ($\perp$) to the orbifold line. Such a decomposition is
introduced because the orbifold line will constitute the new boundary for the
open orbifolds such that the fields straightforwardly decompose into tangential
and normal components with respect to the boundary. The results can thereby be
applied to more general orbifolds. For the torus and its orbifolds the parallel
components correspond to the $y$ direction in the plane and the perpendicular
ones to the $x$ direction, such that a generic vector $\vb{X}$ has complex
components $X_z=X_\perp+iX_\parallel$ and $X_\bz=X_\perp-iX_\parallel$.

\subsection{\label{subsec.fieldtransf}Field Transformations}

Under the $\Omega CT$ symmetry the fields transform as
\be
\ba{cccc}
\Omega CT:&t      &\longmapsto&1-t\vspace{.1 cm}\\
          &x_\perp&\longmapsto&-x_\perp\vspace{.1 cm}\\
          &\Lambda&\longmapsto&\Lambda\vspace{.1 cm}\\
          &A_0    &\longmapsto&-A_0\vspace{.1 cm}\\
          &A_\perp &\longmapsto&-A_\perp\vspace{.1 cm}\\
          &A_\parallel&\longmapsto&A_\parallel\vspace{.1 cm}\\
          &\partial_iE^i &\longmapsto&-\partial_iE^i\vspace{.1 cm}\\
          &B      &\longmapsto&-B\vspace{.1 cm}\\
          &Q_{m,n}    &\longmapsto&Q_{m,n} \ .
\ea
\lb{PCT}
\ee
The orbifold obtained from the quotient under this symmetry corresponds to
\textit{Neumann} boundary conditions on the string fields which restricts the
charge spectrum to only Kaluza-Klein modes $Q_{m,0}=m$. For $\Sigma=T^2$ and
$\tau=i\tau_2$ this induces a two-dimensional open conformal field theory
living on an annulus $C^2$ at the boundaries $\Sigma_0$ and $\Sigma_1$.

For the $\Omega T$ symmetry the fields transform as
\be
\ba{cccc}
\Omega T:&t      &\longmapsto&1-t\vspace{.1 cm}\\
         &x_\perp&\longmapsto&-x_\perp\vspace{.1 cm}\\
         &\Lambda&\longmapsto&-\Lambda\vspace{.1 cm}\\
         &A_0    &\longmapsto&A_0\vspace{.1 cm}\\
         &A_\perp&\longmapsto&A_\perp\vspace{.1 cm}\\
         &A_\parallel&\longmapsto&-A_\parallel\vspace{.1 cm}\\
         &\partial_iE^i &\longmapsto&\partial_iE^i\vspace{.1 cm}\\
         &B      &\longmapsto&B\vspace{.1 cm}\\
         &Q_{m,n}   &\longmapsto&-Q_{m,n} \ .
\ea
\lb{T2}
\ee
The orbifold from the quotient under this symmetry corresponds to
\textit{Dirichlet} boundary conditions on the string fields which restricts the
charge spectrum to only winding modes $Q_{0,n}=kn/4$. Again for $\Sigma=T^2$
and $\tau=i\tau_2$ this induces a two-dimensional open conformal field theory
living on an annulus $C^2$ at the boundaries $\Sigma_0$ and $\Sigma_1$.

Regarding the orbifold operations from the perspective of the full
three-dimensional gauge theory, we see that a single new boundary is created at
the $T$ orbifold fixed point $t=1/2$ as depicted in fig.~\ref{fig.orbi}. The
new boundary $\Sigma_{\frac12}$ may now have a boundary corresponding to the
orbifold fixed points $x_\perp=0$. Furthermore, the presence of
$\Sigma_{\frac12}$ constrains the remaining gauge theory. This means that the
boundary conditions imposed on the fields at the fixed time slice
$\Sigma_{\frac12}$ will effectively propagate through the bulk and constrain
the conformal field theory living on $\Sigma_0\equiv\Sigma_1$ as well. More
specifically, we are working with the boundary conditions of the open conformal
field theories in the twisted sectors of the corresponding Hilbert space.

\fig{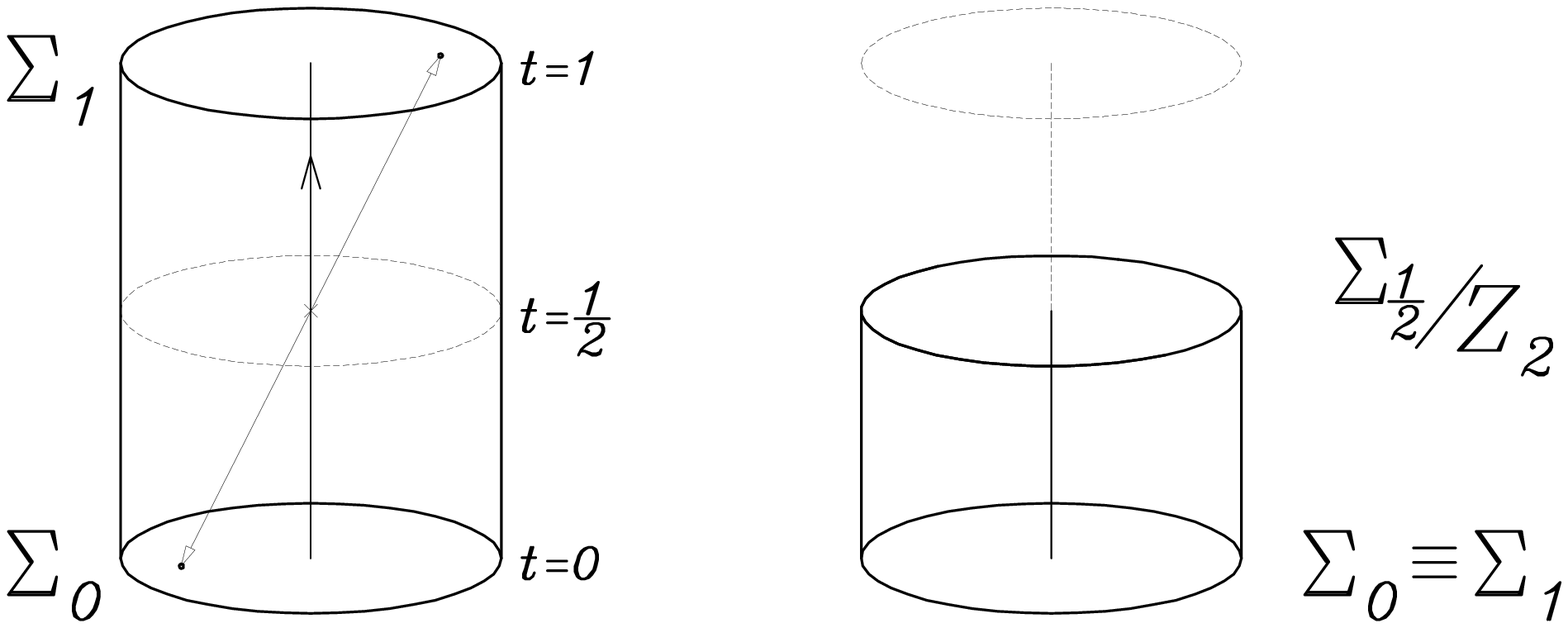}{Orbifold of topologically massive gauge theory. The worldsheet
$\Sigma_{\frac{1}{2}}$ only feels the discrete symmetries $\Omega C$ and
$\Omega$ which each generate the cyclic group ${\mathbb{Z}}_2$.}{fig.orbi}

For later use, we summarize the orbifold identifications on the fixed point
surface $\Sigma_{\frac12}$ and its boundary $\partial\Sigma_{\frac12}$:
\be
\ba{cc}
\mathbf{\Omega T}&\mathbf{\Omega CT}\\[1mm]
\ba{cc}
\underline{\Sigma_{\frac12}}&\underline{\partial\Sigma_{\frac12}}\\[4mm]
A_z\equiv A_\bz&A_z\in {\mathbb{R}}\\[1mm]
A_\perp\equiv A_\perp&A_\perp\neq 0\\[1mm]
A_\parallel\equiv -A_\parallel&A_\parallel=0
\ea&~~~~
\ba{cc}
\underline{\Sigma_{\frac12}}&\underline{\partial\Sigma_{\frac12}}\\[2mm]
A_z\equiv-A_\bz&A_z\in i\,{\mathbb{R}}\\[1mm]
A_\perp\equiv -A_\perp&A_\perp=0\\[1mm]
A_\parallel\equiv A_\parallel&A_\parallel\neq 0 \ .
\ea
\ea
\label{orbid}
\ee

\subsection{\label{subsec.orbpart}Orbifold Partition Functions}

To deal with the new boundary $\Sigma_{\frac12}$ in terms of wavefunctions and
partition functions, we insert into (\ref{ZV}) a complete set of states at time
$t=1/2$ in order to check compatibility with the orbifolding,
\be
Z_\Psi=\sum_{\lambda\in({\mathbb{Z}}_{pq})^g}~
\sum_{\lambda'\in({\mathbb{Z}}_{pq})^g}\,\Bigl\langle\Psi_\lambda(0)\Bigm|
\Psi_{{\lambda}'}(\mbox{$\frac12$})\Bigr\rangle\Bigl\langle
\Psi_{{\lambda}'}(\mbox{$\frac12$})\Bigm|\Psi_{\bar{\lambda}}(1)
\Bigr\rangle \ .
\ee
We now need to determine which wavefunctions can exist at the orbifold fixed
point $t=1/2$. For this, let us consider the Wilson line $W_{Q,{\cal C}}$
corresponding to an electric charge $Q$ propagating along some oriented path
$\cal C$ in the underlying three-manifold. It transforms under the $\Omega CT$
and $\Omega T$ symmetries as $W_{Q,{\cal C}}\mapsto W_{Q,-{\cal
C}}$~\cite{TM_10,TM_16}, so that under the orbifold operation the
identification
\be
W_{Q,{\cal C}}\equiv W_{Q,-{\cal C}}
\lb{W-W}
\ee
must hold for arbitrary contours $\cal C$. Let us consider, in particular, a
path $\cal C$ which lies entirely within the Riemann surface
$\Sigma_{\frac12}$, i.e. at fixed time $t=1/2$. In order that such Wilson lines
be compatible with the orbifold construction they must have vanishing charge,
$Q=0$, or else~(\r{W-W}) is violated. We now recall that the boundary
wavefunctions are generated by Wilson lines that live in the surface $\Sigma$
(see~(\r{Psi-W}) and~\cite{BN_1,LR_1}). Since the only such Wilson line which
is compatible with the orbifold operation is $W_{Q=0,{\cal C}}$, the only
wavefunction that can live at $\Sigma_{\frac12}$ is
$\left|\Psi_{\lambda=0}(\frac12)\right>=\left|1\right>$.

The orbifold partition function is therefore given by
\be
\ba{lll}
\displaystyle Z_{\rm orb}&=&\displaystyle
\sum_{\lambda\in({\mathbb{Z}}_{pq})^g}\,\Bigl\langle\Psi_\lambda(0)
\Bigm|\pi^\Omega\Bigm|\Psi_{\bar{\lambda}}(1)\Bigr\rangle\\[2mm]
&=&\displaystyle
\sum_{\lambda\in({\mathbb{Z}}_{pq})^g}\,\Bigl\langle\Psi^{\rm orb}_\lambda
\Bigm|1\Bigr\rangle \ ,
\ea
\label{piOmega}
\ee
where
\be
\pi^\Omega=\frac12\,\Bigl(\id+\Omega\Bigr)
\label{piOmega1}
\ee
is the orthogonal projection onto parity-even states, i.e. states which are
invariant under the interchange of left and right moving worldsheet modes, and
$\Psi^{\rm orb}_\lambda=\Psi_\lambda(0)\equiv\Psi_{\bar{\lambda}}(1)$. Thus by
inserting a projection operator we \textit{transform} a correlator between two
wavefunctions into the correlator between one wavefunction and unity. This
result is depicted schematically in fig.~\r{figcorr}. It is well-known that
there is a one-to-one correspondence between the Hilbert space of physical
states of the three-dimensional topological quantum field theory and the space
of characters of the induced two-dimensional conformal field theories. The
results obtained here are completely consistent with the expressions for
conformal field theory partition functions in terms of sums over characters of
the conformal group. For closed oriented worldsheets it is a bilinear form in
the characters $\chi^{~}_\lambda$,
\be
\ba{lll}
\displaystyle Z_\Psi&=&\displaystyle\langle\Psi_0|\Psi_1\rangle\\[2mm]
&=&\displaystyle
\sum_{\lambda\in({\mathbb{Z}}_{pq})^g}\overline{\chi^{~}_{\bar{\lambda}}}
{}~\chi^{~}_{\lambda} \ ,
\ea
\ee
while for open or unoriented worldsheets it is simply linear in the characters,
\be
\ba{lll}
\displaystyle Z_{\rm orb}&=&\displaystyle
\left\langle\left.\Psi_{0}^{\rm orb}\right|1\right
\rangle\\[2mm]&=&\displaystyle
\sum_{\lambda\in({\mathbb{Z}}_{pq})^g}\chi_\lambda^{~} \ .
\ea
\ee
We will see some explicit examples later on.

\figbig{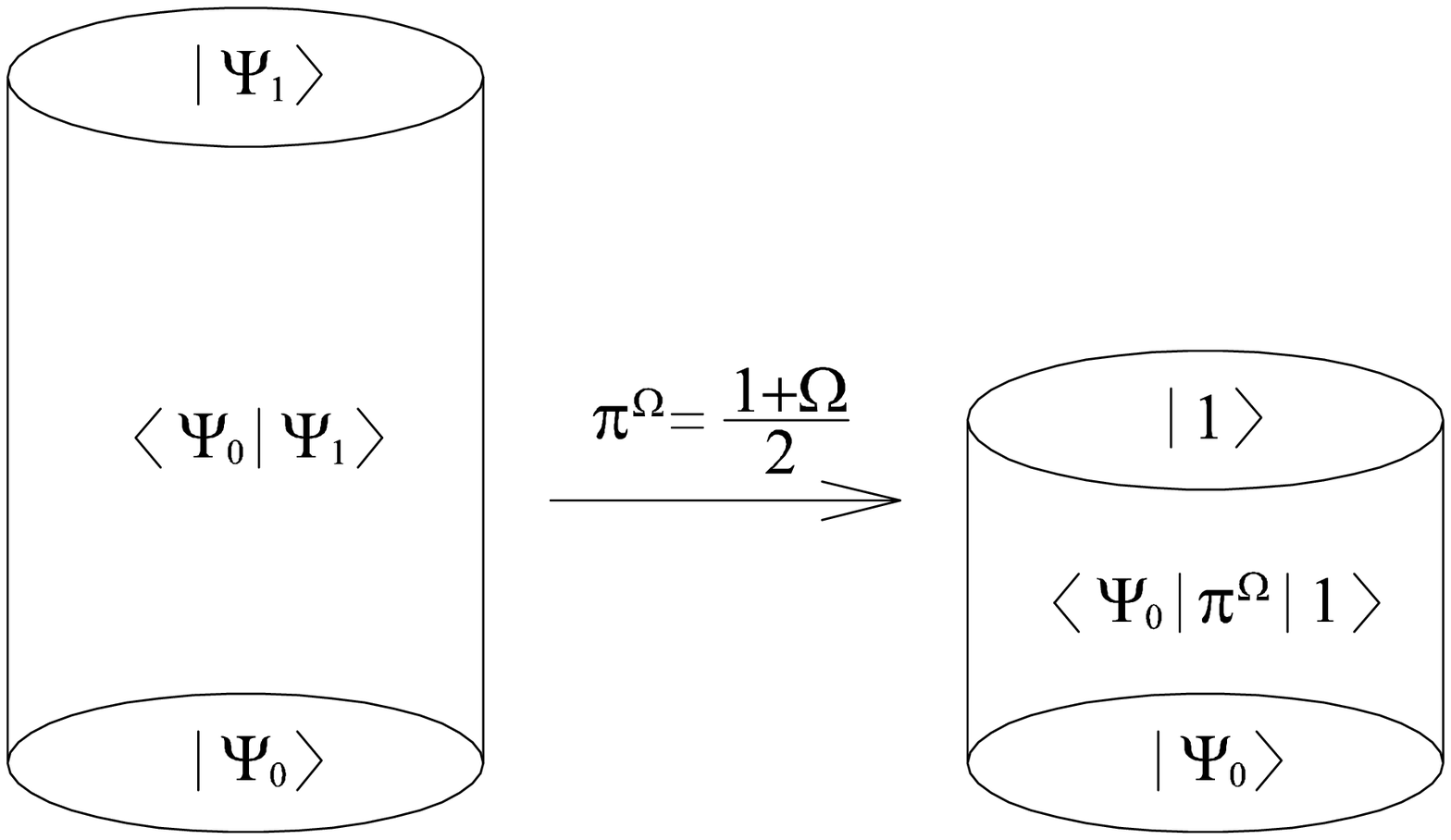}{The resulting orbifold construction in terms of
wavefunction correlators for the gauged parity symmetry $\Omega$.}{figcorr}

\setcounter{equation}{0}

\section{\lb{sec.wave}Chern-Simons Wavefunctions and Amplitudes}

In this section we will construct the quantum wavefunctions of the
three-dimensional gauge theory and use them to study the partition function
appropriate to the topological membrane.

\subsection{\label{subsec.ground}Schr\"odinger Wavefunctionals}

Let us examine the topologically massive gauge theory in the functional
Schr\"odinger picture whereby the physical states are the wavefunctionals
$\Psi[A_i]$~\cite{AFC_1}. By using the canonical quantum commutators
(\ref{com_EB}) and the representation
\be
\Pi^i=-i\,\sqrt h\,\frac\delta{\delta A_i} \ ,
\ee
we impose Gauss' law (\ref{GaussEB}) as a constraint equation on the
wavefunctionals, which takes the form
\be
\left[\partial_\bz\left(-i\,\frac{\delta}{\delta A_\bz}+\frac{k}{8\pi}\,
\tilde\epsilon^{\,z\bz}\,A_z\right)+\partial_z\left(-i\,\frac{\delta}
{\delta A_z}-\frac{k}{8\pi}\,\tilde\epsilon^{\,z\bz}\,A_\bz\right)+
\frac{k}{4\pi}\,\tilde\epsilon^{\,z\bz}\,F_{z\bz}+J^0
\right]\Psi[A_z,A_\bz]=0 \ ,
\ee
where the two-dimensional antisymmetric tensor $\tilde\epsilon$ is induced from
the bulk by $\tilde\epsilon^{\,ij}=\epsilon^{0ij}/\sqrt{-g}$. By applying the
Hamiltonian (\ref{Ham}) to such physical wavefunctions in this polarization, we
find that the stationary states satisfy the functional Schr\"odinger equation
\be
\ba{ll}
&\displaystyle\int\limits_\Sigma d^2z~\sqrt h\,\left\{-\frac\gamma2\,h_{z\bz}\,
\left(-i\,\frac\delta{\delta A_\bz}-\frac k{8\pi}\,\tilde\epsilon^{\,z\bz}
\,A_z\right)\left(-i\,\frac\delta{\delta A_z}+\frac k{8\pi}\,\tilde
\epsilon^{\,z\bz}\,A_\bz\right)\right.\\[2mm]&~~~~~~~~~~\displaystyle
+\left.\frac1{8\gamma}\,\left(\tilde
\epsilon^{\,z\bz}\,F_{z\bz}\right)^2-A_zJ^\bz-A_\bz J^z
\right\}\Psi[A_z,A_\bz]~=~{\cal E}\,\Psi[A_z,A_\bz] \ ,
\ea
\label{Schreq}
\ee
where $\cal E$ is the energy of the state, and we have appropriately normal
ordered the Hamiltonian density. Since the gauge constraint commutes with the
Hamiltonian, these two equations can be consistently solved.

Because of the appearence of the magnetic field in (\ref{U}), in topologically
massive gauge theory the physical states are not gauge invariant. Instead, by
integrating the infinitesimal gauge constraint one finds that the gauge
symmetry is represented projectively on wavefunctionals as
\be
U_\Lambda\Psi[A_i]=\e^{i\alpha[A_i,\Lambda]}\,\Psi[A_i+\partial_i\Lambda] \ ,
\label{projrep}
\ee
where the projective phase is given by the one-cocycle
\be
\alpha[A_i,\Lambda]=\frac k{8\pi}\,\int\limits_\Sigma d^2z~\epsilon^{ij}\,
A_i\,\partial_j\Lambda \ .
\label{cocycle}
\ee
To separate out the gauge invariant part, one integrates the corresponding
cocycle condition such that the physical states decompose into three factors as
\be
\Psi[A_z,A_\bz]=\exp\left\{-\frac{ik}{8\pi}\,\int\limits_\Sigma d^2z~
\sqrt{h}\,\tilde{\epsilon}^{\,z\bz}\,A_zA_\bz\right\}~\psi[A_z]~\Phi[B] \ ,
\lb{Psi}
\ee
where $B$ is the magnetic field.

The factor $\Phi[B]$ is the gauge-invariant solution of Gauss' law for the pure
Maxwell theory ($k=0$),
\be
\left[\partial_\bz\,\frac{\delta}{\delta A_\bz}+\partial_z\,
\frac{\delta}{\delta A_z}\right]\Phi=0 \ .
\ee
If the fields have non-trivial magnetic charge $\int_\Sigma d^2z~\sqrt h\,B
\neq 0$, then the wavefunction $\Phi[B]$ vanishes~\cite{AFC_1}. This result is
simply the statement that there is overall charge conservation on the closed
surface $\Sigma$. Of course, locally non-zero magnetic field distributions are
possible. The component $\psi[A_z]$ obeys the functional Schr\"odinger equation
for the source-coupled, pure Chern-Simons theory ($\gamma\to\infty$),
\be
\left[\partial_z\,\frac{\delta}{\delta A_z}-\frac{i k}{4\pi}\,
\tilde{\epsilon}^{\,z\bz}\,\partial_\bz A_z-i\,J^0
\right]\psi[A_z]=0 \ .
\label{CSsols}\ee

In the following we will only be interested in the ground state of the theory.
This corresponds to a projection onto the lowest Landau level of the quantum
spectrum, which is attained in the topological limit $\gamma\to\infty$ (the
mass gap between Landau levels is proportional to $\gamma$). In this case,
$\Phi=1$ and we recover the wavefunctions of pure Chern-Simons gauge theory.
These solutions correspond to configurations with weak magnetic field,
$\tilde\epsilon^{\,z\bz}\,F_{z\bz}\simeq 0$, and also weak external flux. The
stationary states are the eigenfunctions of the first term on the left-hand
side of (\ref{Schreq}). In particular, the vacuum state (${\cal E}=0$) is
determined by the zero mode equation
\be
\left[\frac\delta{\delta A_z}-\frac k{8\pi}\,A_\bz\right]\Psi[A_z,A_\bz]=0 \ ,
\label{0mode}
\ee
where we have fixed a complex structure on $\Sigma$ determined by the
worldsheet metric whereby $\tilde\epsilon^{\,z\bz}=i$. Furthermore, by using
(\ref{CSsols}) and (\ref{Psi}) with $\Phi=1$, one finds that it obeys the
equation
\be
\left[\partial_\bz\,\frac{\delta}{\delta A_\bz}+\frac{k}{8 \pi}\,
\partial_\bz A_z-\frac{k}{4\pi}\,F_{\bz z}-J^0\right]\Psi[A_z,A_\bz]=0 \ .
\label{WW}
\ee
A solution of (\ref{0mode}) and (\r{WW}), which is compatible with the gauged
WZNW construction for an abelian gauge group, is given by a path integral over
an auxilliary worldsheet field $\varphi$ as~\cite{W_1}
\be
\Psi[A_z,A_\bz]=\int D \mu(\varphi)~\exp\left\{\,\int\limits_{\Sigma}d^2z~
\left[\frac{k}{8\pi}\,\Bigl(A_\bz A_z - 2 A_\bz\,\partial_z\varphi+
\partial_\bz\varphi\,\partial_z\varphi\Bigr)-J^0\,\varphi\right]\right\} \ ,
\label{WWS1}
\ee
where the functional integration measure is given by
\be
 D \mu(\varphi)= D \varphi~\sqrt{{\cal A}_\Sigma}~
\delta\left(\,\int\limits_\Sigma d^2z~\varphi(\vb{z})\,\phi^{~}_0\right)
\label{varphimeas}
\ee
with ${\cal A}_\Sigma=\int_\Sigma d^2z~\sqrt h$ the area of $\Sigma$.

That (\ref{WWS1}) solves (\ref{WW}) follows from noting that the integrand of
the left-hand side of (\ref{WW}) can then be written as a total functional
$\varphi$-derivative of the exponential in (\ref{WWS1}), and thereby integrated
by parts to zero in $\varphi$ space. The factors in the integration measure
(\ref{varphimeas}) remove the zero modes of the field $\varphi$ on the Riemann
surface $\Sigma$, which is required for a well-defined functional integral
because, by charge conservation on a compact space, the exponential in
(\ref{WWS1}) is independent of them. Here $\phi^{~}_0=1/\sqrt{{\cal A}_\Sigma}$
is the normalized zero mode eigenfunction of the scalar Laplace operator
$\nabla_\Sigma^2$ on $\Sigma$, and the delta-function in (\ref{varphimeas}),
whose argument is the coefficient of $\phi^{~}_0$ in an arbitrary field
configuration, projects out the zero mode integration from the measure
$D\varphi$. The worldsheet area factor is included to make the overall
combination dimensionless. A more convenient way to use this measure is to
change variables from $\varphi$ to its worldsheet derivatives and compute the
Jacobian to get
\be
 D \mu(\varphi)= D (\partial_z
\varphi)~ D (\partial_\bz\varphi)~\sqrt{\frac{{\cal A}_\Sigma}
{{\det}'\,\nabla_\Sigma^2}} \ ,
\label{measchange}
\ee
where the prime on the determinant means that zero-modes are excluded. We will
see that the wavefunctions (\ref{WWS1}) are the \textit{building blocks} of the
boundary theories, in that by inserting such states on the boundaries they
\textit{act} as boundary conditions and effectively select the boundary world.
Through the fields $\varphi$ we are introducing new degrees of freedom on the
boundaries which are absolutely necessary for the consistency of the full
three-dimensional theory (on a manifold with boundary) as a well-defined gauge
theory.

\subsection{\label{subsec.inner}Inner Products}

In~\cite{BN_1,W_1} the partition function is defined through the inner
product $\langle\Psi_1|\Psi_2\rangle$ of wavefunctions at a fixed
slice $\Sigma$ of a three-manifold $M$. One chooses a Heegaard
splitting $M=M_1\,\#^{~}_\Sigma\,M_2$ of the manifold such that the
boundaries of the two components are identified, up to a
homeomorphism, as $\Sigma=\partial M_1\cong\partial M_2$. However, the
present topological membrane case (for which $M=[0,1]\times\Sigma$) is
somewhat different since the partition function actually represents a
quantum correlation between two different states, living at times
$t=0$ and $t=1$, corresponding to the two boundaries $\Sigma_0$ and
$\Sigma_1$. We can either interpret it as the amplitude for time
evolution of a state inserted at one boundary to the other boundary,
or as the correlator of two states defined for every time slice whereby each
state decays on different boundaries. This latter interpretation is the
approach that we will follow in this paper.

Let us consider two states of the form~(\r{WWS1}) inserted at the boundaries
$\Sigma_0$ and $\Sigma_1$. The partition function for the topological limit of
the system is then the correlation function
\be
\ba{lll}
\displaystyle Z_\Psi&=&\displaystyle\left<\Psi_0|\Psi_1\right>\\[2mm]
&=&\displaystyle
\int D \mu^{~}_G(A_z,A_\bz)~\e^{iS_{\rm TMGT}[A]}~\overline{{\Psi}_0
[A_z,A_\bz]}\,\Psi_1[A_z,A_\bz] \ ,
\ea
\lb{Z01}
\ee
where $D \mu^{~}_G(A_z,A_\bz)$ is the gauge fixed functional integration
measure for the gauge fields whose precise form will be given in the next
section. The wavefunctional $\Psi_1$ is given by~(\r{WWS1}), while
$\overline{{\Psi}_0}$ is given by
\be
\overline{{\Psi}_0[A_z,A_\bz]}=\int D \mu(\varphi)~
\exp\left\{-\int\limits_{\Sigma}d^2z~\left[\frac{k}{8\pi}\,
\Bigl(A_\bz A_z-2 A_z\,\partial_\bz\varphi+\partial_\bz\varphi\,
\partial_z\varphi\Bigr)-J^0\,\varphi\right]\right\} \ ,
\label{WWS0}
\ee
where the overall minus sign comes from the relative change in orientation
between the surfaces $\Sigma_0$ and $\Sigma_1$. As we will now demonstrate, the
importance of the insertions of these boundary wavefunctions lies in the fact
that they constrain the theory, assuring that the path integral~(\r{Z}) is both
gauge invariant and has a classical extremum. The quantum field theory would
not be consistent without them.

Under a local gauge transformation
\be
\delta_\Lambda A_i=\partial_i\Lambda
\ee
the bulk exponential factor in the path integral (\ref{Z01}) changes according
to
\be
\ba{lll}
\delta_\Lambda S_{\rm TMGT}[A]&=&\displaystyle\int\limits_{\Sigma_1}d^2z~\left[
\frac{k}{8\pi}\,\Bigl(A_z\,\partial_\bz\Lambda-A_\bz\,\partial_z\Lambda
\Bigr)+J^0\,\Lambda\right]\\[2mm]&&\displaystyle-\,
\int\limits_{\Sigma_0}d^2z~\left[
\frac{k}{8\pi}\,\Bigl(A_\bz\,\partial_z\Lambda-A_z\,\partial_\bz\Lambda
\Bigr)+J^0\,\Lambda\right] \ .
\ea
\lb{Sgauge1}
\ee
It is in (\ref{Sgauge1}) that the gauge non-invariance of the theory resides
due to the existence of boundaries, and this leads to chiral WZNW degrees of
freedom on the worldsheet $\Sigma$. On the other hand, because of
(\ref{projrep}) the wavefunction (\ref{WWS1}) at the boundary $\Sigma_1$
transforms as
\be
\delta_\Lambda\Psi_1[A_z,A_\bz]=\int D \mu(\varphi)~
\exp\left\{\frac{k}{8\pi}\,\int\limits_{\Sigma_1}d^2z~
\Bigl[A_\bz\,\partial_z\Lambda + A_z\,\partial_\bz\Lambda+\partial_z\Lambda
\,\partial_\bz\Lambda- 2\,\partial_z\varphi\,\partial_\bz\Lambda
\Bigr]\right\} \ .
\lb{Sgauge2}
\ee
Combining all the factors in~(\r{Sgauge1}) and~(\r{Sgauge2}), we find that the
$\Lambda$-dependent terms of the gauge variation of (\ref{Z01}) can be absorbed
into the total gauge transformed wavefunction at $\Sigma_1$ given by
\be
\ba{lll}
\Psi^{\Lambda}_1[A_z,A_\bz]&=&\displaystyle\int D\mu(\varphi)~\exp\left\{~
\int\limits_{\Sigma_1}d^2z~\left[\frac{k}{8\pi}\,\Bigl(A_\bz A_z - 2 A_\bz\,
\partial_z(\varphi-\Lambda)\Bigr.\right.\right.\\[2mm]&&\displaystyle+
\Biggl.\biggl.\Bigl.\partial_z(\varphi-\Lambda)\,
\partial_\bz(\varphi-\Lambda)\Bigr)-J^0\,(\varphi-\Lambda)\biggr]\Biggr\} \ .
\ea
\lb{Sgauge3}
\ee
An analogous result holds for the wavefunction $\overline{\Psi_0}$ at the
boundary $\Sigma_0$. The gauge parameter $\Lambda$ is now easily eliminated by
redefining the auxilliary field corresponding to the degree of freedom on the
boundary as $\varphi\mapsto\varphi+\Lambda$, such that
$\Psi_1^\Lambda[A_z,A_\bz]=\Psi_1[A_z,A_\bz]$ and
$\overline{\Psi_0^\Lambda[A_z,A_\bz]}=\overline{\Psi_0[A_z,A_\bz]}$. This
redefinition does not change the functional integration measure
$D\mu(\varphi)$, and thus $\delta_\Lambda Z_\Psi=1$. It is in this elegant way,
by inserting seemingly \textit{ad hoc} new degrees of freedom, that we manage
to ensure gauge invariance of the full theory. In other words, gauge
transformations necessarily induce new degrees of freedom on the boundaries,
whose dynamics are governed by the chiral $U(1)$ gauged WZNW
models~(\ref{WWS1}).

Next we show that the boundary wavefunctions ensure that the field theory has a
classical extremum. For this, let us consider the case $J^\mu=0$ for clarity.
An infinitesimal variation of the gauge fields induces a variation of the
action of the form
\be
\delta S_{\rm TMGT}[A,J=0]=\frac{k}{8\pi}\,\left(~\int\limits_{\Sigma_1}d^2z~
\epsilon^{ij}A_i\,\delta A_j-\int\limits_{\Sigma_0}d^2z~
\epsilon^{ij}A_i\,\delta A_j\right) \ .
\lb{dSdM}
\ee
{}From (\ref{dSdM}) it follows that the action $S_{\rm TMGT}$ has no
well-defined classical extremum, because the boundary terms prevent one from
setting $\delta S_{\rm TMGT}=0$. Let us denote the fields $\varphi$
corresponding to the worldsheets $\Sigma_0$ and $\Sigma_1$ by $\varphi^{~}_0$
and $\varphi^{~}_1$, respectively. Using the change of variables
(\ref{measchange}), we may integrate the wavefunction product that appears in
the partition function (\ref{Z01}) to get\footnote{Here and in the following we
drop all (infinite) numerical constants via suitable normalizations of the
functional integration measures. Other infinite constants are evaluated using
standard zeta-function regularization.}
\be
\ba{lll}
Z_\Psi[J=0]&=&\displaystyle
\frac{k\,{\cal A}_\Sigma}{{\det}'\,\nabla_\Sigma^2}\,\int
D\mu^{~}_G(A_z,A_\bz)~
\e^{iS_{\rm TMGT}[A]}\\[2mm]&&\displaystyle\times\,\int\!\!\!\int
D(\partial_\bz\varphi^{~}_1)~D(\partial_z\varphi^{~}_0)~
\delta\left(2A_\bz\bigm|_{\Sigma_1}-\partial_\bz\varphi^{~}_1\right)
\,\delta\left(2A_z\bigm|_{\Sigma_0}-\partial_z
\varphi^{~}_0\right)\\[2mm]&&\displaystyle\times\,
\exp\left\{\frac k{8\pi}\,\left(~\int\limits_{\Sigma_1}d^2z~A_\bz A_z-
\int\limits_{\Sigma_0}d^2z~A_\bz A_z\right)\right\} \ .
\ea
\label{ZPsi}
\ee
In this way the gauge field $A_\bz$ is fixed on one boundary while $A_z$ is
fixed on the other boundary such that $\left.\delta
A_\bz\right|_{\Sigma_1}=\left.\delta A_z\right|_{\Sigma_0}=0$. The overall
boundary variation in (\ref{ZPsi}) containing $\delta A_i$ vanishes since the
term (\ref{dSdM}) which comes from the bulk cancels exactly with the one coming
from the boundary exponential term in the wavefunctions. In this way the gauge
field theory has a well-defined classical limit. Notice that the fields
$\varphi$ living on $\Sigma_0$ correspond to holomorphic degrees of freedom
while the ones living on $\Sigma_1$ correspond to antiholomorphic degrees of
freedom. This method of constructing the two chiral conformal field theories on
the boundary is thereby equivalent to that described
in~\cite{TM_05,TM_15} (see also~\cite{W_1,TM_17}) by fixing
the fields on the boundary and adding extra terms to the boundary action.

\setcounter{equation}{0}

\section{\lb{sec.path}Orbifold Wavefunctions and Amplitudes}

In this section we will generalize the analysis of the previous section to the
orbifold constructions of section~\ref{sec.orbi}. In order to simplify some
arguments which follow we will work with an effective field theory obtained by
integrating the bulk dependence of the gauge fields in the path integral. In
order to do so we need to gauge fix the functional integral by the standard
Faddeev-Popov procedure. For this, we decompose the gauge field $A_i$ according
to a representative of its gauge orbit as
\be
A_i=\bar{A}_i+\partial_i\Lambda \ ,
\ee
where $\bar{A}_i$ is the gauge-fixed field and $\Lambda$ is an arbitrary
real-valued gauge parameter function. Then the measure appearing in (\ref{Z01})
is defined by
\be
D\mu^{~}_G(A_z,A_\bz)=D\bar{A}_z~D\bar{A}_\bz~D\mu(\Lambda)~\sqrt{
\frac{{\det}'\,\nabla_\Sigma^2}{{\cal
A}_\Sigma}}\,\delta\Bigl(G(A)\Bigr)\,\det\left[
\frac{\delta G(A)}{\delta \Lambda}\right] \ ,
\ee
where we have used (\ref{measchange}). Here $G(A)=0$ is the gauge-fixing
condition which will be specified later on. As we have seen, the un-orbifolded
theory is gauge invariant due to the insertion of new degrees of freedom
$\varphi$ on the boundary. Therefore the path integral does not depend on
$\Lambda$ and its integration gives an overall (infinite) constant. However, as
we will see, this may not be the case once one orbifolds the theory. This is
due to the orbifold singularities which play the role of a new boundary of
$\Sigma$, and in a very similar way to the process described
in~\cite{TM_05} a new degree of freedom emerges on the boundary. Put
differently, when $\partial\Sigma\neq\emptyset$, one must generalize the
analysis of the previous section by solving also the Gauss' law constraint on
the boundary of the worldsheet, which appears as the second term in
(\ref{gauss}). As we will show, it will actually be responsible for setting the
boundary conditions of the conformal field theory, so that this process picks
the correct boundary conditions required by the theory in order to restore
gauge invariance.

Assuming that the measure $D\bar{A}_z~D\bar{A}_\bz$ can be factorized into a
bulk integration times a boundary integration, we find that the path integral
factorizes into a product
\be
Z=Z_{\mathrm{bulk}}\,Z_{\Psi} \ ,
\ee
where $Z_{\Psi}$ is the partition function (\ref{Z01}) containing the boundary
wavefunctions. It is evident from the analysis of the previous sections that
these two factors are completely decoupled. We may then integrate over the bulk
gauge fields and effectively normalize $Z_{\mathrm{bulk}}=1$. Generally, it
yields a topological invariant of the underlying framed three-manifold which is
constructed from the Chern-Simons invariant, the gravitational
$\eta$-invariant, and the Ray-Singer analytic torsion. The boundary integration
over $\bar A_i$ also factorizes into two contributions, one for each boundary
component.

\subsection{\label{subsec.orbwave}Orbifold Path Integral}

Within the framework of the path integral approach to the topological membrane,
we can now proceed to orbifold the theory. In order to do so we need
information about the wavefunctions for every time $t$, not just for $t=0$ and
$t=1$, and in particular at the new boundary at $t=1/2$ obtained from the
orbifolding. For this, we consider an extension of the exponential integral
over $\Sigma$ defining the previous wavefunctions~(\r{WWS1}) and~(\r{WWS0}) to
the full three-dimensional manifold $[0,1]\times\Sigma$. The most efficient way
to achieve this is to insert two generic time-dependent functions $f_0(t)$ and
$f_1(t)$ with the boundary conditions
\be
\ba{rrl}
f_0(0)~=~-f_1(1)&=&-1 \ , \\[2mm]f_0(1)~=~f_1(0)&=&0 \ ,
\ea
\lb{fs}
\ee
and rewrite the wavefunction product appearing in the path integral (\ref{Z01})
as
\be
\ba{ll}
&\displaystyle\overline{\Psi_0\left[{A}_z,A_\bz\right]}\,
\Psi_1\left[{A}_z,A_\bz\right]\\[2mm]&\displaystyle~~~~=
|{\cal N}|^2\,\int D\mu(\varphi)~\exp\left\{\,\int
\limits_0^1dt~\int\limits_\Sigma d^2z~\frac\partial{\partial t}\left[\left(
\frac{k}{8\pi}\,\Bigl({A}_\bz{A}_z+\partial_\bz\,\varphi\,\partial_z
\varphi\Bigr)-J^0\,\varphi\right)(f_0+f_1)\right.
\right.\\[3mm]&\displaystyle~~~~~~~~-\left.\left.
\frac k{4\pi}\,\Bigl({A}_z\,\partial_z\varphi\,f_0 -{A}_\bz\,\partial_\bz
\varphi\,f_1\Bigr)\right]\right\} \ ,
\ea
\lb{Wa}
\ee
where we have extended the scalar field $ \varphi $, previously inserted only
at the boundaries through the functional integrals present in the definitions
(\ref{WWS1}) and (\ref{WWS0}), to the whole three-manifold $[0,1]\times\Sigma$.
We have also introduced a normalization constant $\cal N$ in the definition of
the wavefunctions, which will be fixed later on by certain orthonormality
requirements. A possible profile for the temporal functions $f_0$ and
$f_1$ is depicted in fig.~\r{figfs}. Their introduction allows the
implementation of the desired program by extending the correlator of two
wavefunctions (defined at each time slice) to the full three-dimensional bulk
theory.

\figbig{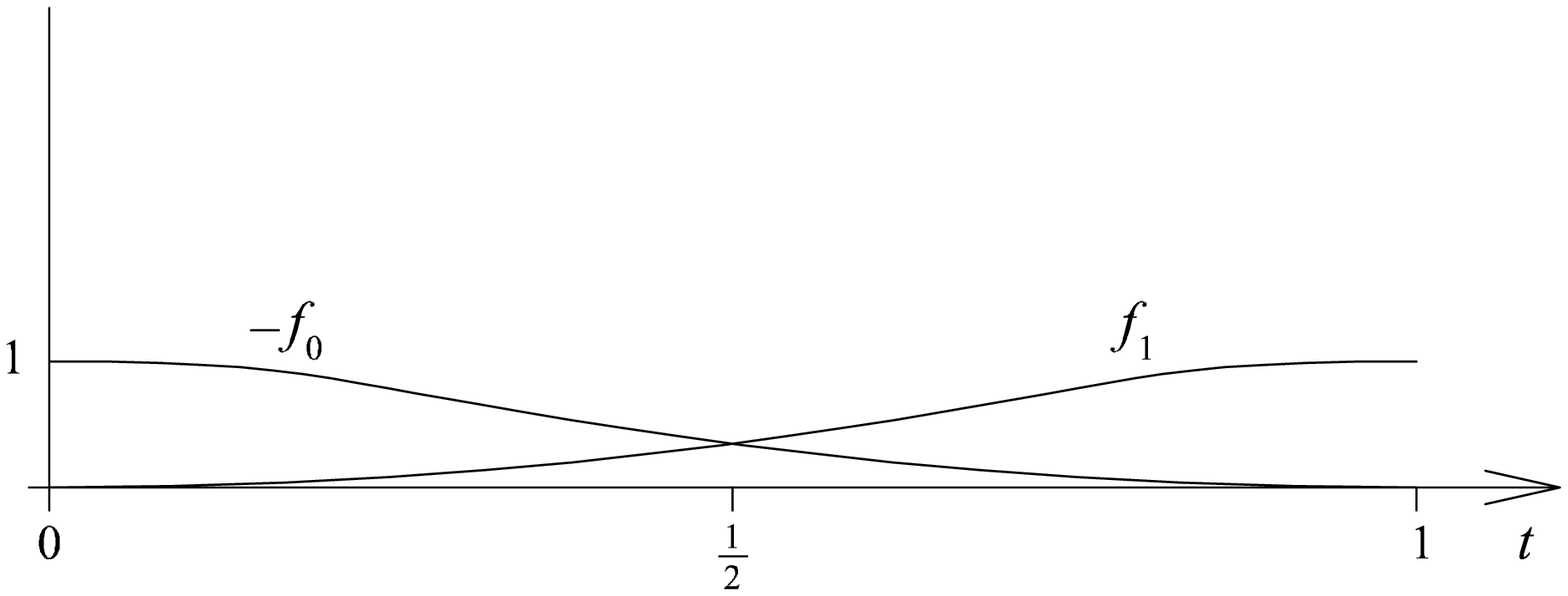}{Profile of the decay and growth of the temporal functions
$-f_0(t)$ and $f_1(t)$.}{figfs}

Let us consider the orbifold by the discrete symmetries $\Omega CT$ and
$\Omega T$, taking, for consistency, the identifications of the field
$\varphi$ to be of the same kind as the gauge parameter $\Lambda$ in
section~\ref{subsec.fieldtransf}. Since $t=1/2$ is a fixed point of the
identification, we split the time integral over $[0,1]$ into two pieces over
$[0,\frac12]$ and $(\frac12,1]$. Rewriting~(\r{Wa}) with this splitting and
performing the required identifications we conclude that, for compatibility
with the discrete symmetries, the time-dependent functions must be related by
\be
f_0(1-t)=-f_1(t) \ .
\ee
Putting everything together we find that both parts of the temporal integration
give the same contribution. Furthermore, since $f_0(\frac12)=-f_1(\frac12)$,
the product (\r{Wa}) reduces under the orbifold operation to
\be
\ba{ll}
&\displaystyle\overline{\Psi_{0}^{\rm orb}\left[A_z,A_\bz\right]}
\,\Psi_{\frac{1}{2}}^{\rm orb}\left[A_z,A_\bz\right]\\[2mm]&
\displaystyle~~~~=\overline{\cal N}\,{\cal N}_{\rm orb}
\,\int D\mu(\varphi)~\exp\left\{2\int
\limits_0^{1/2}dt~\int\limits_\Sigma d^2z~\frac\partial{\partial t}
\left[\left(
\frac{k}{8\pi}\,\Bigl({A}_\bz{A}_z+\partial_\bz\,\varphi\,\partial_z
\varphi\Bigr)-J^0\,\varphi\right)\right.
\right.\\[2mm]&\displaystyle~~~~~~~~\times\left.\left.(f_0+f_1)-
\frac k{4\pi}\,\Bigl({A}_z\,\partial_z\varphi\,f_0 -{A}_\bz\,\partial_\bz
\varphi\,f_1\Bigr)\right]\right\} \ ,
\ea
\lb{Wext}
\ee
where we have naturally defined two new \textit{orbifold} wavefunctions at
times $t=0$ and $t=1/2$ by
\be
\ba{rll}
\displaystyle\overline{\Psi_{0}^{\rm orb}\left[{A}_z,A_\bz
\right]}&=&\displaystyle\overline{\cal N}
\,\int D\mu(\varphi^{~}_0)~\exp\left\{-\frac{k}{4\pi}\,\int\limits_{\Sigma}d^2z
{}~\Bigl[{A}_\bz{A}_z - 2{A}_\bz\,\partial_z \varphi^{~}_0  +
\partial_\bz \varphi^{~}_0\,\partial_z \varphi^{~}_0\Bigr]\right\} \ , \\[2mm]
\displaystyle\Psi_{\frac{1}{2}}^{\rm orb}\left[{A}_z,A_\bz\right]&=&
\displaystyle{\cal N}_{\rm orb}\,\int D\mu\left(\varphi^{~}_{\frac12}\right)~
\exp\left\{\frac{k\,f_0(\mbox{$\frac12$})}
{2\pi}\,\int\limits_{\Sigma^O}d^2z~\left[{A}_z\,\partial_\bz
\varphi^{~}_{\frac12}-{A}_\bz\,\partial_z \varphi^{~}_{\frac12}
\right]\right\} \ ,
\ea
\lb{Wo12}
\ee
with $\Sigma^O=\Sigma/{\mathbb{Z}}_2[O]$ and ${\mathbb{Z}}_2[O]$ the cyclic
group generated by the order two symmetry $O=\Omega$ or $O=\Omega C$ that we
have chosen to quotient the theory by. Here and in the following we set the
external charge density $J^0=0$ on the $t=0$ surface $\Sigma_0$. We have also
introduced a new normalization constant ${\cal N}_{\rm orb}$ for the
wavefunction at $t=1/2$, which will be computed later on from the original
normalization $\cal N$.

Let us now consider the orbifold path integral
\be
\ba{lll}
\displaystyle Z_{\rm orb}&=&\displaystyle
\left<\Psi_{0}^{\rm orb}\left|\Psi_{\frac{1}{2}}^{\rm orb}
\right.\right>\\[2mm]&=&\displaystyle
\int D\mu^{~}_G(A _z,A_\bz)~\e^{2iS^{\rm orb}_{\rm TMGT}[A]}~
\overline{\Psi_{0}^{\rm orb}[A_z,A_\bz]}\,
\Psi_{\frac{1}{2}}^{\rm orb}[A_z,A_\bz] \ ,
\ea
\ee
where $S_{\rm TMGT}^{\rm orb}[A]$ is the topologically massive gauge theory
action (\ref{STMGT}) for the orbifold of the three-dimensional theory, and the
orbifold wavefunctions are given by~(\r{Wo12}). Here we have identified the
actions~\cite{H_1}
\be
2S_{\rm TMGT}^{\rm orb}\left[A^{\rm orb}\right]=S_{\rm TMGT}[A]
\ee
with $A$ regarded as the extension of the gauge field $A^{\rm orb}$ from
$\Bigl([0,1]\times\Sigma\Bigr)/{\mathbb{Z}}_2$ to the covering cylinder. To
ease notation in the following we will not explicitly write the superscript orb
on the orbifold fields. In the bulk of the worldsheet $\Sigma^O$ (neglecting
boundary contributions over $\partial\Sigma^O$), this orbifold quantum field
theory is gauge invariant. Under the bulk transformation~(\r{Sgauge1}) of the
action, one may check explicitly that it is enough to shift the auxilliary
field $ \varphi $ on the two boundaries as
\be
\ba{lll}
\varphi^{~}_0&\longmapsto&\varphi^{~}_0-\Lambda \ , \\[2mm]
\varphi^{~}_{\frac12}&\longmapsto&\displaystyle\varphi^{~}_{\frac12}-
\frac{\Lambda}{2f_0(\mbox{$\frac12$})}
\ea
\ee
in order to restore the gauge independence of all the factors under a bulk
gauge transformation. What is particularly interesting here is that the
wavefunction $\Psi_{\frac{1}{2}}^{\rm orb}[A_z,A_\bz]$ which emerges at the
orbifold fixed point yields exactly the factor necessary to cancel the gauge
dependent contribution coming from~(\r{Sgauge1}).

We can also rewrite the wavefunctions in terms of the normal and
parallel components defined in section~\ref{sec.orbi} to get
\be
\Psi_{\frac{1}{2}}^{\rm orb}[A_z,A_\bz]={\cal N}_{\rm orb}\, \int
D\mu(\varphi)~\exp\left\{\frac{ik\,f_0(\mbox{$\frac12$})}{\pi}\,\int
\limits_{\Sigma^O}d^2z\left[A_\parallel\,\partial_\perp \varphi
-A_\perp\,\partial_\parallel \varphi \right]\right\} \ .  \lb{Wo12pp}
\ee
Looking back at~(\r{orbid}) we see a naive indication that the
$\Omega T$ orbifold, containing only the term $\partial_{\parallel}
\varphi $ at the boundary, imposes \textit{Dirichlet} boundary
conditions, while the $\Omega CT$ orbifold, for which only the factor
with $\partial_{\perp} \varphi $ is present at the boundary, imposes
\textit{Neumann} boundary conditions. We will derive these results
formally in the next subsection. Therefore, the emergence of the wavefunction
$\Psi_{\frac{1}{2}}^{\rm orb}[A_z,A_\bz]$ correctly imposes the
anticipated boundary conditions at the orbifold line. In the next
subsection we will also see that this wavefunction corresponds to the
identity operator of the boundary conformal field theory. Note that in
the cases where $\partial\Sigma^O=\emptyset$, we would get instead
\textit{twisted} and \textit{untwisted} closed string sectors for the
$\Omega T$ and $\Omega CT$ orbifolds, respectively, as studied
in~\cite{TM_16}.

However, as mentioned before, this is not the whole story, because if
$\Sigma^O$ has a boundary, then the gauge fixed path integral acquires a gauge
dependent piece and is given explicitly by
\be
\ba{lll}
\displaystyle Z_{\rm orb}&=&\displaystyle
\sqrt{\frac{{\det}'\,\nabla^2_{\Sigma^O}}
{{\cal A}_{\Sigma^O}}}\,\int D\bar{A}_z~D\bar A_\bz~
\e^{2iS_{\rm TMGT}^{\rm orb}[\bar A]}~
\overline{\Psi_{0}^{\rm orb}\left[\bar{A}_z,\bar A_\bz\right]}\,
\Psi_{\frac{1}{2}}^{\rm orb}\left[\bar A_z,\bar A_\bz\right]
\\[2mm]&&\displaystyle\times\,\int\!\!\!\int D\mu(\Lambda)~
D\mu\left(\tilde\varphi\right)~\delta\Bigl(G(A)\Bigr)\,\det\left[
\frac{\delta G(A)}{\delta\Lambda}\right]\,
\exp\left\{\frac{ik\,f_0(\mbox{$\frac12$})}\pi
\,\oint\limits_{\partial\Sigma^O}\Lambda\,\partial_\perp
\tilde\varphi\right\} \ ,
\ea
\lb{Zbb}
\ee
where $\tilde\varphi$ is the restriction of the auxilliary field
$\varphi^{~}_{\frac12}$ to the boundary $\partial\Sigma^O$. We now need to
analyse carefully the orbifold identifications and check compatibility. For
this, we consider the Hodge decompositions of the gauge fields $A_i$ at each
time slice, which in the holomorphic polarization read
\be
\ba{lll}
A_z&=&a_z + \partial_z\overline{\phi} \ , \\[2mm]
A_\bz&=&a_\bz + \partial_\bz\phi \ ,
\ea
\lb{HD}
\ee
where
\be
\phi=\phi^{~}_R+i\,\phi^{~}_I
\ee
is a complex scalar field and $a_i$ is a
harmonic one-form with
\be
\partial_za_\bz\pm\partial_\bz a_z=0 \ .
\ee
The harmonic one-forms have quantized periods $a/2\pi\in
H^1(\Sigma,{\mathbb{Z}})$. Extending the identifications~(\r{orbid}) on
$\Sigma_{\frac12}=\Sigma^O$ to the degrees of freedom $a_i$ and $\phi$ and to
the gauge parameter $\Lambda$ one gets:
\be
\ba{cc}
\mathbf{\Omega T}&\mathbf{\Omega CT}\\[2mm]
\ba{cc}
\underline{\Sigma^\Omega}&
\underline{\partial\Sigma^\Omega}\\[2mm]
a_z\equiv a_\bz&a_z\in {\mathbb{R}}\\[2mm]
a_\parallel\equiv -a_\parallel&a_\parallel=0\\[2mm]
a_\perp\equiv a_\perp&a_\perp\neq 0\\[2mm]
\phi^{~}_R\equiv-\phi^{~}_R&\phi^{~}_R=0\\[2mm]
\phi^{~}_I\equiv\phi^{~}_I&\phi^{~}_I\neq 0\\[2mm]
\Lambda\equiv-\Lambda&\Lambda=0
\ea&~~~~
\ba{cc}
\underline{\Sigma^{\Omega C}}&
\underline{\partial\Sigma^{\Omega C}}\\[2mm]
a_z\equiv-a_\bz&a_z\in i\,{\mathbb{R}}\\[2mm]
a_\parallel\equiv a_\parallel&a_\parallel\neq 0\\[2mm]
a_\perp\equiv-a_\perp&a_\perp=0\\[2mm]
\phi^{~}_R\equiv\phi^{~}_R&\phi^{~}_R\neq 0\\[2mm]
\phi^{~}_I\equiv-\phi^{~}_I&\phi^{~}_I=0\\[2mm]
\Lambda\equiv\Lambda&\Lambda\neq 0 \ .
\ea
\ea
\lb{orbidHD}
\ee

\subsection{\label{subsec.bcs}Boundary Conditions}

To analyse the partition function (\ref{Zbb}), we need to first specify a gauge
condition. For this, we note that a gauge transformation only shifts the field
$\phi^{~}_R\mapsto\phi^{~}_R+\Lambda$. So a natural choice is to impose the
gauge constraint
\be
G(A)=\phi+\overline{\phi}=0 \ ,
\ee
so that the gauge fixed fields are
\be
\ba{lll}
\bar{A}_z&=&a_z - i\,\partial_z\phi^{~}_I \ , \\[2mm]
\bar{A}_\bz&=&a_\bz + i\,\partial_\bz\phi^{~}_I
\ea
\lb{HDgf}
\ee
while the scalar field $\phi^{~}_R$ has been integrated out and is no longer
present in the path integral. Substituting this decomposition into the orbifold
wavefunction $\Psi_{\frac{1}{2}}^{\rm orb}$ at the new boundary at $t=1/2$
yields
\be
\Psi_{\frac{1}{2}}^{\rm orb}\left[\bar A_z,\bar A_\bz\right]
={\cal N}_{\rm orb}\,\int D\mu(\varphi)~
\exp\left\{\frac{ik\,f_0(\mbox{$\frac12$})}{2\pi}\,
\int\limits_{\Sigma^O}d^2z~
\left[\epsilon^{ij}\,a_i\,\partial_j \varphi +2\,\sqrt h\,h^{ij}\,
\partial_i\phi^{~}_I\,\partial_j \varphi \right]\right\} \ .
\lb{Wo12HD}
\ee
Integrating by parts and using the fact that $a_i$ is a harmonic form then
gives the path integral
\be
\ba{lll}
\displaystyle Z_{\rm orb}&=&\displaystyle
\sqrt{\frac{{\det}'\,\nabla^2_{\Sigma^O}}
{{\cal A}_{\Sigma^O}}}\,{\cal N}_{\rm orb}\,
\int D\bar{A}_z~D\bar A_\bz~\e^{2iS_{\rm TMGT}^{\rm orb}[\bar A]}~
\overline{\Psi_{0}^{\rm orb}\left[\bar A_z,\bar A_\bz\right]}\\[2mm]
&&\displaystyle\times\,\int\!\!\!\int D\mu(\Lambda)~D\mu(\varphi)~
\exp\left\{\frac{ik\,f_0(\mbox{$\frac12$})}{\pi}\,
\int\limits_{\Sigma^O}d^2z~\phi^{~}_I\,
\nabla_{\Sigma^O}^2 \varphi\right.\\[2mm]&&\displaystyle+\left.
i\,\oint\limits_{\partial\Sigma^O}\left[\frac{k\,f_0(\mbox{$\frac12$})}{\pi}\,
\Bigl(a_\perp\,\varphi+\Lambda\,\partial_\perp \varphi +\phi^{~}_I\,
\partial_\parallel \varphi\Bigr)-2\,\phi^{~}_I\,J^\parallel\right]\right\} \ ,
\ea
\label{Zorbexpl}
\ee
where the measure for functional integration over the gauge fields is given by
\be
D\bar A_z~D\bar A_\bz=Da_z~Da_\bz~
D\mu(\phi^{~}_I)~\sqrt{\frac{{\det}'\,\nabla_\Sigma^2}{{\cal A}_\Sigma}} \ .
\ee
The boundary source term in (\ref{Zorbexpl}) arises from the minimal coupling
in the action upon integrating by parts, and using the continuity equation
(\ref{Jmuconserved}) along with charge conservation on closed surfaces. With
our choice of time-dependence for the conserved currents $J^\mu$, this is the
only place that source terms remain.

One sees immediately that the integration over $\phi^{~}_I$ yields the Dirac
delta-function constraint $\nabla_{\Sigma^O}^2 \varphi =0$ in the bulk of
$\Sigma^O$. This is expected because it corresponds to the equation of motion
of the free chiral boson in the boundary conformal field theory. When
$\partial\Sigma^O=\emptyset$, this constraint is the only remnant of the
insertion of the wavefunction $\Psi_{\frac12}^{\rm orb}$. For the cases where
$\partial\Sigma^O\neq\emptyset$, there are two types of wavefunctions
$\Psi_{\frac12}^{\rm orb}$ corresponding to the two kinds of orbifolds under
$\Omega T$ and $\Omega CT$. As we will now explicitly demonstrate, they
correspond to the two possible conformal boundary conditions,
\textit{Dirichlet} and \textit{Neumann}. More precisely, we will show that the
orbifold path integral contains Dirac delta-functions correspondingly as:
\be
\ba{rrl}
{\mathrm{Dirichlet}} \ \ (\Omega T)&:&
\delta(\partial_\parallel\varphi)~\delta(\varphi-Y) \ , \\[2mm]
{\mathrm{Neumann}} \ \ (\Omega CT)&:&\delta(\partial_\perp\varphi) \ ,
\ea
\lb{Psio12}
\ee
where $Y$ is some fixed function on $\partial\Sigma^O$. To do so, we must
carefully analyse the field configurations compatible with the identifications
in~(\r{orbidHD}).

For the $\Omega T$ quotient, $a_\perp\neq 0$, $\Lambda=0$ and
$\phi^{~}_I\neq 0$, and so the integrations over both $\phi^{~}_I$ and
$a_\perp$ on $\partial\Sigma^\Omega$ are present and yield the
respective Dirac delta-function constraints $\partial_\parallel
\varphi =2\pi\,J^\parallel/k\,f_0(\frac12)$ and $ \varphi =0$. These two
conditions are compatible with each other and correspond to \textit{Dirichlet}
boundary conditions, the first constraint specifying that the field $\varphi$
is fixed at the boundary and the second one specifying what function
it is fixed to. Indeed, by fixing the components of the external current
parallel to the boundary line $\partial\Sigma^O$ to be
\be
J^\parallel=\frac{k\,f_0(\frac12)}{2\pi}\,\partial_\parallel Y \ ,
\label{Jparallel}
\ee
we can shift the boundary field $\varphi\mapsto\varphi+Y$, and rewrite the
conditions as $\partial_\parallel\varphi=0$ and $\varphi=Y$. The value of the
function on $\partial\Sigma^O$ can thereby be fixed to any desired one by the
appropriate choice of external sources. The function $Y$ appearing in 
(\ref{Jparallel}) may in fact be interpreted as the collective coordinate of a
D-brane, and hence the minimal coupling to conserved currents allows the
construction of D-branes in topological membrane theory~\cite{progress}.
Note that in this case the path integral does not depend on the gauge
parameter $\Lambda$, so that these types
of orbifolds do not break the gauge symmetries.

For the $\Omega CT$ orbifolds, $a_\perp= 0$, $\Lambda\neq 0$ and
$\phi^{~}_I= 0$, and so only the integration over $\Lambda$ on
$\partial\Sigma^{\Omega C}$ is present, yielding the Dirac
delta-function constraint $\partial_\perp \varphi =0$. It corresponds
to \textit{Neumann} boundary conditions.  Alternatively, we can think
of this requirement as stating that the only boundary conditions
compatible with these types of orbifolds are Neumann boundary
conditions, because they are the only ones which are gauge invariant.
This constraint could also have been deduced by using the Gauss law
constraint (\ref{gauss}) over $\partial\Sigma^{\Omega C}$ to solve for
the orbifold boundary dependence of the wavefunctionals analogously to
the construction of section~\ref{sec.wave}. The key identity which
allows for this is the zero-mode equation (\ref{0mode}) satisfied by the
vacuum wavefunctionals (\ref{WWS1}). In this way it is possible to
deal, in a completely analogous way, with the functional equation for
the normal and longitudinal components on $\partial\Sigma^{\Omega C}$,
and one easily checks that the gauge invariant solution thereby
selects the value \be f_0(\mbox{$\frac12$})=\frac12 \ .  \ee

\subsection{\label{subsec.loccontr}Normalizations}

The main conclusion of the previous subsection is that the insertion of the
orbifold wavefunction at $t=1/2$ merely serves to constrain the bulk dynamics
of the WZNW field $\varphi$ on $\Sigma^O$ with the appropriate boundary
conditions. After integration over $\varphi$, there is no further field
dependence from this boundary, and the quantum state on $\Sigma^O$ is formally
unity $|1\rangle$, in agreement with our earlier expectations. Similarly,
integrating out the local parts $\varphi$ and $\phi^{~}_I$ of the wavefunction
at $t=0$ produces some overall determinant factors, along with the integrations
over the harmonic forms $a_i$. The latter quantities transform under large
gauge transformations, thereby encoding the topological nature of the solutions
and the correspondence between the wavefunctions and conformal blocks. They
will be dealt with in the next section. In this subsection we will carefully
tabulate the determinant factors that arise from integration over the local
components of the fields. They will incorporate part of the moduli dependence
of the resulting states.

The key ingredient in the computation is the fixing of the normalization
constant $\cal N$. The canonical choice is to demand that the norms of an
orthogonal basis of wavefunctions in the unorbifolded theory be equal to
unity, implemented by the functional integration
\be
1=\int D\mu^{~}_G(A_z,A_\bz)~\overline{\Psi_0[A_z,A_\bz]}\,
\Psi_0[A_z,A_\bz] \ .
\label{fnintnorm}
\ee
The corresponding partition function is then a vacuum-to-vacuum amplitude and
it leads to a well-defined inner product on the quantum Hilbert space of
physical states. The way to correctly define this object is by identifying
(gluing) the two boundaries $\Sigma_0$ and $\Sigma_1$. The explicit
computation can be found in~\cite{TM_14,TM_17} while the correct relative
boundary conditions are given in~\cite{TM_05,TM_15}. One finds that
the fields $\phi^{~}_I$ coming from the Hodge decomposition of the
gauge field must be identified at the two boundaries at $t=0$ and $t=1$, while
the boundary conformal field theory degrees of freedom are combined into the
single scalar field $\varphi^{~}_d = \varphi^{~}_0- \varphi^{~}_1$. Note that
while $\varphi_0^{~}$ and $\varphi^{~}_1$ are chiral worldsheet fields, their
difference $\varphi^{~}_d$ is not. The Lagrangian for the field $\varphi^{~}_d$
is
proportional to $k\,\partial_z\varphi^{~}_d\,\partial_\bz\varphi^{~}_d$ and its
functional integration measure is $D\mu(\varphi^{~}_d)$. The path integral over
this field thereby produces precisely the same determinant factor that appears
in (\ref{measchange}), times a factor $\sqrt k$ involving the Chern-Simons
coefficient. The integration over the gauge degree of freedom $\phi^{~}_I$
produces no such determinant factors and yields the overall constant $\sqrt k$,
because its functional integration measure is
$D(\partial_z\phi^{~}_I)~D(\partial_\bz\phi^{~}_I)$ and it only appears in the
path integral through its worldsheet derivatives. As mentioned at the beginning
of this section, the same is true of the integration over the gauge parameters
$\Lambda$. Thus, to properly normalize the wavefunctionals of
section~\ref{sec.wave}, we will take the normalization constant
\be
{\cal N}=\left(\frac{{\det}'\,\nabla_\Sigma^2}{k^2\,
{\cal A}_\Sigma}\right)^{1/4} \ .
\label{C}
\ee
Integrating out all the local fields in the path integral (\ref{ZPsi}) in this
way produces an effective inner product $(\cdot,\cdot)_a$ depending solely on
the harmonic gauge fields $a_i$ on the two boundaries $\Sigma_0$ and
$\Sigma_1$. This inner product will be worked out explicitly in the next
section. Incorporating the normalization factor (\ref{C}) into (\ref{ZPsi}) we
then arrive at
\be
Z_\Psi=\sqrt{\frac{{\cal
A}_\Sigma}{{\det}'\,\nabla_\Sigma^2}}~(\Psi_0,\Psi_1)_a \ .
\label{ZPsieff}
\ee
In (\ref{ZPsieff}) the wavefunctions are obtained by integrating out all local
field dependences.

We can now easily work out the normalization constant ${\cal N}_{\rm orb}$ of
the orbifold wavefunctions from the un-orbifolded one (\ref{C}). The crucial
observation is that $\cal N$ is defined through the path integral with
action $k\int_\Sigma d^2z~\partial_z\varphi^{~}_d\,\partial_\bz\varphi^{~}_d$.
Under the orbifold operations the integrand of this action is invariant but the
orientation of the worldsheet $\Sigma$ reverses due to the transformation
$z\mapsto-\bz$, i.e. this action vanishes on the resulting orbifolded surface
$\Sigma^O$. Note that this fact was used already in the construction of the
orbifold wavefunction at $t=1/2$ in (\ref{Wo12}). In other words, the
determinant of the scalar Laplacian on $\Sigma^O$ with the given orbifold
boundary conditions is equal to unity. We thereby conclude that
\be
{\cal N}_{\rm orb}=1 \ .
\ee
Completely analogous arguments establish that the functional integrals over
$\phi^{~}_I$ and $\Lambda$ at $t=1/2$ contribute only the usual $\sqrt k$
constants to the orbifold partition function (\ref{Zorbexpl}). The only
non-trivial local field dependence that remains comes from the integral over
the $t=0$
wavefunction, through its normalization constant (\ref{C}) and its dependence
on $\varphi^{~}_0$ which yields the determinant factor in (\ref{measchange}).
Remembering that the wavefunction at $t=1/2$ is unity, the orbifold path
integral is thereby given through the properly normalized topological inner
product
\be
Z_{\rm orb}=\left(\frac{{\cal A}_\Sigma}{{\det}'\,\nabla_\Sigma^2}\right)^{1/4}
\,\left(\Psi_0^{\rm orb},1\right)_a \ .
\label{Zorbeff}
\ee
Note that the orbifold state $\overline{\Psi_0^{\rm orb}}$ corresponding to
$\Sigma_0$ is identical to that in (\ref{WWS0}), except that the argument of
its exponential is twice as large. Again, in (\ref{Zorbeff}) it is understood
that the wavefunction depends only on the harmonic one-forms $a_i$.

\setcounter{equation}{0}

\section{\label{sec.character}Character Expansions}

In this section we will finally come to the main objective of this
paper.  We will complete the evaluation of the orbifold partition
functions of the previous section, using them to construct modular
invariants for the induced conformal field theories. For this, we will
evaluate the path integrals over the harmonic components $a_i$ of the
gauge fields, assuming that the the external currents $J^\mu$ are
cohomologically trivial, as is appropriate for the vacuum amplitudes. As an
explicit example, we will look at various toroidal orbifolds and show that the
topological membrane formalism yields the anticipated results in these cases.

\subsection{\label{subsec.effwave}Effective Topological Wavefunctions}

We now proceed to determine the dependence of the wavefunctions on the
harmonic forms $a_i$. For this, we work in a holomorphic polarization,
with respect to the chosen complex structure of $\Sigma$ which we take
to be of genus $g$, and decompose the harmonic forms into the
sums~\cite{TM_11}
\be
\ba{lll}
a_z(\bz)&=&\displaystyle\pi i\,\sum_{\ell=1}^ga^\ell\,\left(\Gamma_2^{-1}
\right)^{\ell\ell'}\,\overline{\omega_{\ell'}(z)}
\ , \\[2mm]a_\bz(z)&=&\displaystyle-\pi i\,\sum_{\ell=1}^g\overline{a}^\ell\,
\left(\Gamma_2^{-1}\right)^{\ell\ell'}\,\omega_{\ell'}(z) \ ,
\ea
\ee
where $\omega_\ell=\omega_\ell(z)~dz$, $\ell=1,\dots,g$ form a basis of
holomorphic one-differentials determined by the conditions
\be
\ba{lll}
\displaystyle\oint\limits_{\alpha_\ell}\omega_{\ell'}&=&
\delta^{\ell}_{~\ell'} \ , \\[2mm]
\displaystyle\oint\limits_{\beta^\ell}\omega_{\ell'}&=&\Gamma_{\ell\ell'} \ .
\ea
\ee
Here $\alpha_\ell,\beta^\ell$, $\ell=1,\dots,g$ form a canonical set of closed
contours on the Riemann surface $\Sigma$ which generate its first homology
group $H_1(\Sigma,{\mathbb{Z}})$. The $g\times g$ complex period matrix
$\Gamma$ of the surface is symmetric, and we will decompose it into real and
imaginary parts as $\Gamma=\Gamma_1+i\,\Gamma_2$. Its imaginary part $\Gamma_2$
is a positive definite symmetric matrix. The $\omega_\ell$ span the kernel of
the antiholomorphic derivative $\partial_\bz$ and are normalized as
\be
\int\limits_\Sigma d^2z~\omega_\ell(z)\,\overline{\omega_{\ell'}(z)}=
-2i\,(\Gamma_2)_{\ell\ell'} \ .
\ee

The gauge transformations $\Sigma\to U(1)$ of the theory have integer winding
numbers around homologically non-trivial curves in $\Sigma$. A map which winds
$s^\ell$ times around the primitive cycle $\alpha_\ell$ and $r_\ell$ times
around $\beta^\ell$ determines a large gauge transformation\footnote{Here and
in the following the topological indices $\ell$ are raised and lowered with
the metric $G_{\ell\ell'}=2(\Gamma_2)_{\ell\ell'}$ on the space of
holomorphic harmonic one-forms.}
\be
\ba{lll}
a^\ell&\longmapsto&a^\ell+s^\ell+\Gamma^{\ell\ell'}\,r_{\ell'} \ , \\[2mm]
\overline{a}^\ell&\longmapsto&\overline{a}^\ell+s^\ell+
\overline{\Gamma}^{\ell\ell'}\,r_{\ell'} \ .
\ea
\label{largegt}
\ee
The topological dependence of the Schr\"odinger wavefunctionals may now be
uniquely fixed by requiring that they furnish a projective representation of
the large gauge symmetries. Let us consider a rational Chern-Simons coefficient
$k=2p/q$, with $p$ even. Then the projective representation defines a
semi-periodicity condition on the wavefunctions which is solved by functions
that live in a finite-dimensional vector space. With the holomorphic
parametrization above, we can build an orthogonal basis, with $(pq)^g$ members
($(2pq)^g$ members for $p$ odd), for the wavefunctions in the
form~\cite{TM_11,BN_1,LR_1,W_0}
\be
\Psi_{\lambda}(a,\overline{a};\Gamma)=
\exp\left\{\frac{k\pi}8\,a^\ell\left(
\Gamma_{2}^{-1}\right)_{\ell\ell'}\left(a^{\ell'}-\overline{a}^{\ell'}
\right)\right\}\,\Theta_p\left[\ba{c}\frac{\lambda}{q}\\[1mm]0\ea\right]
\left(\sqrt{2}\,a\left|\frac{2\Gamma}{k}\right.\right) \ ,
\lb{WT2}
\ee
where $\lambda$ is an integer $g$-vector with components in the range
$\lambda^\ell=0,1,\ldots,pq-1$. The genus $g$ {\it modified} Jacobi
theta-functions are defined by
\be
\ba{ll}
&\displaystyle
\Theta_p\left[\ba{c}\frac{\lambda}{q}\\[1mm]0\ea\right]\left(\sqrt{2}\,a\left|
\frac{2\Gamma}{k}\right.\right)\\[2mm]&\displaystyle~~~~~~=~
\sum_{s\in p{\mathbb{Z}}^g}\exp\left\{\frac{2\pi i}
{k}\,\left(s^\ell+\frac{\lambda^\ell}{q}\right)\Gamma_{\ell\ell'}
\left(s^{\ell'}+\frac{\lambda^{\ell'}}{q}\right)+2\,\sqrt{2}\,\pi i\,
a_l\left(s^\ell+\frac{\lambda^\ell}{q}\right)\right\} \ ,
\ea
\label{Theta}
\ee
where by ``modified'' we mean that the integer sums are restricted to
multiples of $p$.

The wavefunctions (\ref{WT2}) are those for the first Landau level on the
Riemann surface $\Sigma$. They carry a $(pq)^g$-dimensional unitary irreducible
representation of the holonomy algebra of the gauge theory in terms of clock
and cyclic shift operators as
\be
\ba{rll}
\displaystyle\exp\left\{i\,\oint\limits_{\alpha_\ell}A\right\}
\Psi_\lambda(a,\overline{a};\Gamma)&=&\e^{2\pi i\lambda^\ell/q}~
\Psi_\lambda(a,\overline{a};\Gamma) \ , \\[2mm]
\displaystyle\exp\left\{i\,
\oint\limits_{\beta^\ell}A\right\}\Psi_{\lambda}
(a,\overline{a};\Gamma)&=&\Psi_{\lambda+q1_\ell\,(\mod pq)}
(a,\overline{a};\Gamma) \ ,
\label{holrep}
\ea
\ee
where $1_\ell$ is the integer $g$-vector with components
$(1_\ell)^{\ell'}=\delta_\ell^{~\ell'}$. As described in
section~\ref{subsec.bdrypart}, the representation (\ref{holrep}) explicitly
realizes the monopole processes which change the charge of the quantum states.
The normalization of the Chern-Simons coefficient, which differs from the
standard ones~\cite{BN_1}, is chosen so that the $U(1)$ charges carried by
Wilson lines belong to the required spectrum~(\r{charge}) of the theory. In
this way, the large gauge transformations (\ref{largegt}) are restricted to
those with $s^\ell,r_\ell=0\,\mod p$. Note that $\lambda^\ell/q$ are  primary
charges and that for the rational values $k=2p/q$ there is one charge
independent monopole process with $\Delta Q=p$. In terms of the induced
conformal field theory these shifts of charge lie within a single
family~\cite{TM_15}. Then the restricted large gauge transformations
(\ref{largegt}) are just the allowed monopole processes which shift charges
inside the same family or conformal block.

Let us now consider the unorbifolded path integral over the topological
components of the gauge fields in (\ref{ZPsieff}). As we did in
section~\ref{subsec.loccontr}, we will start by fixing the normalization of the
effective wavefunctions. For this, we define an inner product between
wavefunctions $\Psi_{\lambda}(a,\overline{a};\Gamma)$ and
$\Psi_{\lambda'}(a,\overline{a};\Gamma)$ by identifying the $a_i$ fields on the
boundaries $\Sigma_0$ and $\Sigma_1$ and integrating out their topological
dependence as
\be
\left(\Psi_{\lambda},\Psi_{\lambda'}\right)_a=\frac1{2^g\,\det\Gamma_2}\,
\int\limits_{{\cal J}(\Sigma)}\,\prod_{\ell=1}^g
da^\ell~d\overline{a}^\ell~\overline{\Psi_{\lambda}(a,\overline{a};\Gamma)}\,
\Psi_{\lambda'}(a,\overline{a};\Gamma) \ ,
\label{topinnerprod}
\ee
where, due to the large gauge invariance of this inner product, the integration
is restricted to the Jacobian variety ${\cal J}(\Sigma)$ of the Riemann surface
$\Sigma$ obtained by reducing the $a,\overline{a}$ phase space to the plaquette
of gauge orbits
\be
\ba{lll}
a^\ell&=&\rho^\ell+\Gamma^{\ell\ell'}\,\gamma^{~}_{\ell'} \ , \\[2mm]
\overline{a}^\ell&=&\rho^\ell+\overline{\Gamma}^{\ell\ell'}\,\gamma^{~}_{\ell'}
\ea
\ee
with $\rho^\ell,\gamma^{~}_\ell\in[0,p]$. This is the inner product that is
inherited from the normalization integral of (\ref{fnintnorm}), and also that
which naturally arises in the geometric quantization of pure Chern-Simons gauge
theory.

As explained in section~\ref{subsec.bdrypart}, one must supplement this vacuum
amplitude with monopole transitions enforcing large gauge invariance, and so we
compute instead
\be
\ba{ll}
&\displaystyle\Bigl(\Psi_\lambda~,~\exp\left\{-i\,\sum_{\ell=1}^g\Delta
Q_\lambda^\ell\,\oint\limits_{\beta^\ell}A\right\}\Psi_{\lambda'}\Bigr)_a\\
[2mm]&\displaystyle~~~~~~=~\sum_{s\in p{\mathbb{Z}}^g}~
\sum_{s'\in p{\mathbb{Z}}^g}~\prod_{\ell=1}^g\,\int
\limits_0^pd\rho^\ell~\exp\left\{2\,\sqrt2\,\pi i\,\rho_\ell\left(s'^\ell
-s^\ell+\frac{\lambda'^\ell}q-\frac{\lambda^\ell}q-\Delta Q_\lambda^\ell
\right)\right\}\\[2mm]
&\displaystyle~~~~~~~~~~~~\times\,\prod_{\ell=1}^g\,\int\limits_0^p
d\gamma^{~}_\ell~\exp\left\{-\frac{k\pi}2\,
\gamma^{~}_\ell\,(\Gamma_2)^{\ell\ell'}
\gamma^{~}_{\ell'}+2\,\sqrt2\,\pi i\,\gamma^\ell\left[\Gamma_{\ell\ell'}
\left(s'^{\ell'}+\frac{\lambda'^{\ell'}}q\right)\right.\right.\\[2mm]
&\displaystyle~~~~~~~~~~~~
-\left.\overline{\Gamma}_{\ell\ell'}\left(s^{\ell'}+\frac{\lambda^{\ell'}}q+
\Delta Q_\lambda^{\ell'}\right)\right]+\frac{2\pi i}k\,\left[\left(s'^\ell+
\frac{\lambda'^\ell}q\right)\Gamma_{\ell\ell'}\left(s'^{\ell'}+
\frac{\lambda'^{\ell'}}q\right)\right.\\[2mm]
&\displaystyle~~~~~~~~~~~~-\left.\left.
\left(s^{\ell}+\frac{\lambda^{\ell}}q+\Delta Q_\lambda^\ell
\right)\overline{\Gamma}_{\ell\ell'}
\left(s^{\ell'}+\frac{\lambda^{\ell'}}q+\Delta Q_\lambda^{\ell'}\right)
\right]\right\} \ .
\ea
\lb{topinnermon}
\ee
The effect of the monopole charges $\Delta Q_\lambda^\ell$ is to shift a charge
$Q_\lambda^\ell$ on one boundary to the charge
$Q_{\bar\lambda}^\ell=Q_\lambda^\ell+\Delta Q_\lambda^\ell$ on the other
boundary. Let us assume a decomposition $Q_\lambda^\ell=\lambda^\ell/q$ of the
charge as in (\ref{lambdadef}), where for each $\ell=1,\dots,g$ the integers
$(m^\ell,n^\ell)$ form a corresponding Bezout pair, i.e. the smallest integers
satisfying this decomposition. Then, as discussed in section~\ref{sec.tmgt},
the opposite charge $Q_{\bar\lambda}^\ell=\bar\lambda^\ell/q$ is given by
(\ref{barlambdadef}) and therefore $\Delta Q_\lambda^\ell=-kn^\ell/2$.

The integration over the variables $\rho^\ell$ in (\ref{topinnermon}) thereby
gives $2^{-g/2}\,\delta^{(g)}_{{\cal
J}(\Sigma)}\left(\frac{\lambda'-\lambda}q-\frac{pn}q\right)$, where the
periodic delta function, invariant under the large gauge transformations
(\ref{largegt}), on the complex $g$-torus ${\cal J}(\Sigma)$ is given by
\be
\delta_{{\cal J}(\Sigma)}^{(2g)}({\mathbf{a}})
=\sum_{s\in p{\mathbb{Z}}^g}~\sum_{r\in p{\mathbb{Z}}^g}
\delta^{(2g)}({\mathbf{a}}+s+\Gamma\cdot r) \ .
\label{deltaper}
\ee
This sets $\lambda'^\ell/q=\bar\lambda^\ell/q\,\mod p$ for each
$\ell=1,\dots,g$. The sum over $s'\in p\,{\mathbb{Z}}^g$ in (\ref{topinnermon})
then enforces the $\mod p$ periodicity in this identification, and upon
completing the square in the resulting terms involving the $\gamma^{~}_\ell$
variables we get
\be
\ba{ll}
&\displaystyle\Bigl(\Psi_\lambda~,~\exp\left\{-i\,\sum_{\ell=1}^g\Delta
Q_\lambda^\ell\,\oint\limits_{\beta^\ell}A\right\}\Psi_{\lambda'}\Bigr)_a\\
[2mm]&\displaystyle~~~~~~=~\frac{\delta_{\lambda',\bar\lambda}}{2^{g/2}}
\,\sum_{s\in p{\mathbb{Z}}^g}~\prod_{\ell=1}^gd\gamma^{~}_\ell~
\exp\left\{-\frac{k\pi}2\,\left[\gamma^\ell+\frac{2\,\sqrt2}k\,
\left(s^\ell+\frac{\bar\lambda^\ell}q\right)\right](\Gamma_2)_{\ell\ell'}
\right.\\[2mm]&\displaystyle~~~~~~~~~~~~\times\left.
\left[\gamma^{\ell'}+\frac{2\,\sqrt2}k\,
\left(s^{\ell'}+\frac{\bar\lambda^{\ell'}}q\right)\right]\right\} \ .
\ea
\label{topinnergamma}
\ee
Shifting variables
$\gamma^\ell\mapsto\gamma^\ell-2\,\sqrt2\,(s^\ell+\bar\lambda^\ell/q)/k$ in
(\ref{topinnergamma}) and using the sum over $s\in p\,{\mathbb{Z}}^g$ to extend
the integration domain from the gauge orbit space to all of ${\mathbb{R}}^g$
yields a Gaussian integral in the $\gamma^{~}_\ell$'s, and we arrive finally at
\be
\Bigl(\Psi_\lambda~,~\exp\left\{-i\,\sum_{\ell=1}^g\Delta
Q_\lambda^\ell\,\oint\limits_{\beta^\ell}A\right\}\Psi_{\lambda'}\Bigr)_a
=\frac1{k^{g/2}\,\sqrt{\det\Gamma_2}}~\delta_{\lambda',\bar\lambda} \ ,
\label{topinnernorm}
\ee
where as before we drop all irrelevant numerical constants.

Using the normalizations (\ref{ZPsieff}) and (\ref{topinnernorm}), we can now
write down the appropriate wavefunctions whose products will generate the
desired partition function. After integrating out the bulk modes, the path
integral (\ref{ZPsi}) decomposes into two independent integrals over the modes
associated with each of the two boundaries. We therefore define the total
effective wavefunctions by integrating out their $a_i$ field dependences
explicitly using the inner product (\ref{topinnerprod}), and we arrive at
\be
\ba{lll}
\Psi_\lambda(\Gamma)&=&\displaystyle\left(\frac{{\cal A}_\Sigma}{{\det}'\left(
\nabla_\Sigma^2\right)\,\det(k\,\Gamma_2)}\right)^{1/4}~\left(1,
\Psi_\lambda\right)_a\\[2mm]&=&\displaystyle\left(\frac{{\cal A}_\Sigma}
{{\det}'\left(\nabla_\Sigma^2\right)\,\det(k\,\Gamma_2)}\right)^{1/4}\,
\sum_{s\in p{\mathbb{Z}}^g}~\prod_{\ell=1}^g~\int\limits_0^p\!\int
\limits_0^pd\rho^\ell~d\gamma^{~}_\ell~\exp\Biggl\{2\pi i\,\Bigl(\rho^\ell+i\,
(\Gamma_2)^{\ell\ell'}\,\gamma^{~}_{\ell'}\Bigr)\Biggr.
\\[2mm]&&\displaystyle\times\left.\left[\frac k8\,
\gamma^{~}_\ell+\sqrt2\,\left(s_\ell+\frac{\lambda_\ell}q\right)\right]+
\frac{2\pi i}k\,\left(s^\ell+\frac{\lambda^\ell}
q\right)\Gamma_{\ell\ell'}\left(s^{\ell'}+\frac{\lambda^{\ell'}}q\right)
\right\}\\[2mm]&=&\displaystyle\left(\frac{{\cal A}_\Sigma}
{{\det}'\left(\nabla_\Sigma^2\right)\,\det(k\,\Gamma_2)}\right)^{1/4}\,
\sum_{s\in p{\mathbb{Z}}^g}\exp\left\{\frac{2\pi i}k\,\left(s^\ell+
\frac{\lambda^\ell}q\right)\Gamma_{\ell\ell'}\left(s^{\ell'}+
\frac{\lambda^{\ell'}}q\right)\right\} \ .
\ea
\label{Psifinal}
\ee

\subsection{\label{subsec.CFTpart}Statistical Sums}

Up to a factor independent of the geometry of the Riemann surface $\Sigma$, the
wavefunctions (\ref{Psifinal}), along with the charge selection rule, naturally
lead to the partition function for the rational conformal field theory of a
compactified free scalar field. In string units the radius of compactification
$R$ is given through the relation $k=2R^2$, so that the conformal field theory
partition function $Z_\Sigma$ is given by
\be
\ba{lll}
Z_\Sigma&=&k^{g/2}~Z_\Psi\\[2mm]&=&\displaystyle k^{g/2}~\sum_{\lambda\in
({\mathbb{Z}}_{pq})^g}\overline{\Psi_{\bar\lambda}(\Gamma)}~\Psi_\lambda
(\Gamma)\\[2mm]&=&\displaystyle
\sqrt{\frac{{\cal A}_\Sigma}{{\det}'\left(\nabla_\Sigma^2\right)
\,\det\Gamma_2}}~\sum_{\lambda\in({\mathbb{Z}}_{pq})^g}~
\sum_{s\in p{\mathbb{Z}}^g}~\sum_{s'\in p{\mathbb{Z}}^g}\exp\left\{
\frac{2\pi i}k\,\left(s^\ell+\frac{\lambda^\ell}{q}\right)
\Gamma_{\ell\ell'}\left(s^{\ell'}+\frac{\lambda^{\ell'}}{q}\right)\right.
\\[2mm]&&\displaystyle-\left.\frac{2\pi i}k\,\left(s^\ell+
\frac{\bar{\lambda}^\ell}{q}\right)\overline{\Gamma}_{\ell\ell'}
\left(s^{\ell'}+\frac{\bar{\lambda}^{\ell'}}{q}\right)\right\} \ .
\ea
\ee
It is now straightforward to compute the orbifold partition functions
given the analysis of the previous subsection. As we have seen before,
for the orbifold theories we have only the wavefunction at the
boundary $\Sigma_0$. Moreover, the charge spectrum is appropriately
truncated~\cite{TM_15}. It is now clear in fact that the integers
$m^\ell$ correspond to the Kaluza-Klein modes of the compactification
while $n^\ell$ correspond to winding numbers. Thus not all of the
wavefunctions $\Psi_\lambda$ obtained above will survive the orbifold
operations. Precisely, only $(pq/2)^g$ of them remain, so that the
dimension of the physical state space is reduced by a factor of
$2^{-g}$. Recall that $p$ is even, so that this counting of states is
coherent. In the case that $p$ is odd the number of linearly
independent states is reduced from $(2pq)^g$ to $(pq)^g$ by the
orbifold operations.

For the $\Omega CT$ symmetries the charge spectrum is truncated to the values
\be
\frac{\lambda^\ell}q=m^\ell \ , ~~ m^\ell=0,\ldots,\frac{pq}2-1 \ ,
\lb{KKm}
\ee
and the harmonic fields obey
$a^\ell=-\overline{a}^\ell=i\,(\Gamma_2)^{\ell\ell'}\,\gamma^{~}_{\ell'}$.
Furthermore, the period matrix is necessarily purely imaginary,
$\Gamma=i\,\Gamma_2$, for compatibility with this orbifold. Following through
the analysis of the previous subsection, in this case there are no $\rho^\ell$
integrations, and together with the doubling of the argument of the exponential
factor in the wavefunction $\Psi_\lambda^{\rm orb}$ we find an overall
normalization factor $2^{-g}$, as expected.

On the other hand, for the $\Omega T$ symmetries the spectrum is truncated as
\be
\frac{\lambda^\ell}q=\frac k4\,n^\ell \ , ~~ n^\ell=0,\ldots,\frac{pq}2-1 \ ,
\lb{wn}
\ee
and in this case
$a^\ell=\overline{a}^\ell=\rho^\ell+(\Gamma_1)^{\ell\ell'}\,
\gamma^{~}_{\ell'}$. There are now no Gaussian $\gamma^{~}_\ell$ integrations
arising, and the gauge orbit integral over the new variables
$\rho^\ell+(\Gamma_1)^{\ell\ell'}\,\gamma^{~}_{\ell'}$ produces an extra
multiplicative factor $2^{-g/2}$. The overall normalization factor obtained in
this case is therefore $2^{-g}\,\sqrt{\det(k\,\Gamma_2)}$.

Explicitly, we have thereby obtained the total orbifold partition functions
\be
\ba{lll}
\displaystyle Z_{\Sigma^O}&=&k^{g/4}~Z_{\rm orb}\\[2mm]&=&
\displaystyle k^{g/4}~\sum_{\lambda\in\bigl({\mathbb{Z}}_{\frac{pq}2}\bigr)^g}
\Psi_\lambda^{\rm orb}(\Gamma) \ ,
\ea
\ee
which for the two types of orbifold constructions under consideration here are
given by
\be
Z_{\Sigma^{\Omega C}}=\frac{1}{2^g}\,
\left(\frac{{\cal A}_\Sigma}{{\det}'\left(\nabla_\Sigma^2\right)
\,\det\Gamma_2}\right)^{1/4}~\sum_{m\in\bigl({\mathbb{Z}}_{\frac{pq}2}\bigr)^g}
{}~\sum_{s\in p{\mathbb{Z}}^g}\exp\left\{\frac{2\pi i}k\,
\left(s^\ell+m^\ell\right)\Gamma_{\ell\ell'}
\left(s^{\ell'}+m^{\ell'}\right)\right\}
\ee
and
\be
Z_{\Sigma^\Omega}=
\frac{k^{g/2}}{2^g}\,\left(\frac{{\cal A}_\Sigma\,\det\Gamma_2}{{\det}'
\,\nabla_\Sigma^2}\right)^{1/4}~\sum_{n\in\bigl(
{\mathbb{Z}}_{\frac{pq}2}\bigr)^g}
{}~\sum_{s\in p{\mathbb{Z}}^g}\exp\left\{\frac{2\pi i}k\,
\left(s^\ell+\frac k4\,n^\ell\right)\Gamma_{\ell\ell'}
\left(s^{\ell'}+\frac k4\,n^{\ell'}\right)\right\} \ .
\ee
These formulas all hold for arbitrary genus $g$. For $g=0$ there are no moduli
and one obtains only trivial determinant factors. The case $g=1$ will be
analysed in detail in the next subsection.

For genus $g\geq2$, the formulas above are invariant under the natural action
of the Siegal modular group $Sp(2g,{\mathbb{Z}})$ on the space of period
matrices. They should be supplemented by the explicit expression for the
determinant of the scalar Laplacian on the hyperbolic Riemann surface $\Sigma$
which is given by~\cite{DHPh}
\be
\ba{lll}
{\det}'\,\nabla_\Sigma^2&=&\displaystyle\frac{{\cal A}_\Sigma\,\det\Gamma_2}
{8\pi^2}~\e^{2S_{\rm L}[h_{z\bz}]}\,\left[\frac{\displaystyle\Theta\left[
\ba{c}0\\[1mm]0\ea\right]\left(\left.
\sum_{\ell=1}^g\,\int\limits_{z_0}^{z_\ell}
\omega-\int\limits_{z_0}^w\omega-\triangle~\right|~\Gamma\right)}{\displaystyle
\det_{\ell,\ell'}\Bigl[\omega_\ell(z_{\ell'})\Bigr]~\sigma(w)}\right.\\[2mm]&&
\displaystyle\times\left.\prod_{\ell=1}^g\frac{\sigma(z_\ell)}{E(z_\ell,w)}
{}~\prod_{\ell'<\ell}E(z_{\ell'},z_\ell)\right]^{4/3} \ ,
\ea
\label{detnablag2}
\ee
where
\be
S_{\rm L}[h_{z\bz}]=\frac1{48\pi}\,\int\limits_\Sigma d^2z~\partial_z
\ln(h_{z\bz})\,\partial_\bz\ln(h_{z\bz})
\label{Liouville}
\ee
is the Liouville action in the conformal gauge, and $z_1,\dots,z_g,w$ are
arbitrary points on $\Sigma$. The ordinary Jacobi theta-functions are given by
\be
\Theta\left[\ba{c}c\\[1mm]d\ea\right](u|\Gamma)=\sum_{s\in{\mathbb{Z}}^g}
\exp\Bigl\{\pi i\left(s^\ell+c^\ell\right)\Gamma_{\ell\ell'}\left(
s^{\ell'}+c^{\ell'}\right)+2\pi i\left(s^\ell+c^\ell\right)\left(
u_\ell+d_\ell\right)\Bigr\}
\label{ordthetafn}
\ee
for $c^\ell,d_\ell\in[0,1]$ and ${\mathbf{u}}\in{\mathbb{C}}^g$, while
\be
\triangle_\ell=\frac12-\frac12\,\Gamma_{\ell\ell}+\sum_{\ell'\neq\ell}
{}~\oint\limits_{\alpha_{\ell'}}dz~\omega_{\ell'}(z)\,\int\limits_{z_0}^z
\omega_\ell
\label{Riemannvec}
\ee
is the vector of Riemann constants which is related to the Riemann divisor
class of $\Sigma$. The prime form is given by
\be
E(z,z_0)=\frac{\displaystyle\Theta\left[\ba{c}1/2\\[1mm]1/2\ea\right]
\left(\left.\,\int\limits_{z_0}^z\omega~\right|~\Gamma\right)}
{\sqrt{h^{~}_E(z)\,h^{~}_E(z_0)}}
\label{primeform}
\ee
with the $\frac12$-differentials
\be
h^{~}_E(z)=\left.\omega^\ell(z)\,\frac\partial{\partial u^\ell}\Theta\left[
\ba{c}1/2\\1/2\ea\right](u|\Gamma)\right|_{u=0} \ .
\label{hEdef}
\ee
The prime form is antisymmetric under interchange of its arguments and behaves
like $z-z_0$ for $z\to z_0$. Finally, the quantities $\sigma(z)$ are
twisted differentials of rank $g/2$ which carry the conformal anomaly, and
which may be defined by choosing a set of points $P_1,\dots,P_g$ on $\Sigma$
and setting their ratios as
\be
\frac{\sigma(z)}{\sigma(w)}=\frac{\displaystyle\Theta\left[\ba{c}0\\[1mm]0
\ea\right]\left(\left.\,\int\limits_{z_0}^z\omega-
\sum_{\ell=1}^g\,\int\limits_{z_0}^{P_\ell}\omega+\triangle~\right|~\Gamma
\right)}{\displaystyle\Theta\left[\ba{c}0\\[1mm]0
\ea\right]\left(\left.\,\int\limits_{z_0}^w\omega-
\sum_{\ell=1}^g\,\int\limits_{z_0}^{P_\ell}\omega+\triangle~\right|~\Gamma
\right)}~\prod_{\ell=1}^g\frac{E(w,P_\ell)}{E(z,P_\ell)} \ .
\label{sigmaratio}
\ee
It follows from the Riemann vanishing theorem that the ratios
(\ref{sigmaratio}) are independent of the choice of points $P_\ell\in\Sigma$,
and moreover that $\sigma(z)$ has no zeroes or poles on $\Sigma$. In this way
we manage to produce explicit expressions for multi-loop string vacuum
amplitudes entirely within the topological membrane approach.

\subsection{\label{subsec.1loop}Examples: One-Loop Amplitudes}

While we have obtained expressions for the partition functions on any Riemann
surface $\Sigma$, it is clear from the previous subsection that in general one
needs to incorporate the gravitational sector of the topological membrane in
order to properly deal with the effective wavefunctions. In this respect it is
illuminating to examine them in some detail in the simplest case of genus
$g=1$, where everything can be worked out explicitly and there is no
gravitational anomaly to contend with. The resulting formulas can be deduced
from the general equations that we have obtained in the previous subsection.
This will serve, among other things, to illustrate explicitly how the
three-dimensional formalism naturally reproduces modular invariants of the
conformal field theories.

\bigskip

\noindent
$\bullet\,\underline{\rm Torus}$

Let us begin with the torus $T^2$. In this case the period matrix may be
represented by the Teichm\"uller modular parameter
$\Gamma=\tau=\tau_1+i\tau_2$, a complex number with imaginary part $\tau_2>0$.
The complex coordinates on $T^2$ are given by the identifications $z\equiv
z+s+r\tau$, for any $s,r\in{\mathbb{Z}}$, in the complex plane, and the
holomorphic harmonic one-form may be taken to be $\omega(z)=1$. The area of the
torus is ${\cal A}_\tau=\tau_2$, and the determinant of the scalar Laplace
operator on $T^2$ is given by
\be
{\det}'\,\nabla_\tau^2=\tau_2\,\Bigl|\eta(\tau)\Bigr|^4 \ ,
\label{dettorus}
\ee
where
\be
\eta(\tau)=\e^{\pi i\tau/12}\,\prod_{r=1}^\infty\left(1-\e^{2\pi ir\tau}\right)
\label{Dedekind}
\ee
is the Dedekind function.

The partition function is therefore given by
\be
\ba{lll}
Z_{T^2}(\tau)&=&\displaystyle\sum_{\lambda=0}^{pq-1}\,\overline{\chi^{~}_{\bar
\lambda}(\tau)}~\chi^{~}_\lambda(\tau)\\[2mm]
&=&\displaystyle\frac{1}{\sqrt{\tau_2}\,\Bigl|\eta(\tau)\Bigr|^2}
\,\sum_{\lambda=0}^{pq-1}~\sum_{s\in p{\mathbb{Z}}}~
\sum_{s'\in p{\mathbb{Z}}}\exp\left\{\frac{2\pi i \tau}k\,
\left(s+\frac{\lambda}{q}\right)^2\right\}\,\exp\left\{-\frac{2\pi
i\overline{\tau}}k\,\left(s+\frac{\bar{\lambda}}{q}\right)^2\right\}
\ea
\label{T2part}
\ee
where the $\chi$'s may be recognized as the simple characters of the relevant
chiral $U(1)$ current algebras extended by chiral vertex operators. They are
related to the gauge theory wavefunctions by
\be
\ba{lll}
\chi^{~}_\lambda(\tau)&=&k^{g/4}~\Psi_\lambda(\tau)\\[2mm]&=&\displaystyle
\frac1{\tau_2^{1/4}\,\eta(\tau)}\,\sum_{s\in p{\mathbb{Z}}}
\exp\left\{\frac{2\pi i\tau}k\,\left(s+\frac\lambda q\right)^2\right\} \ .
\ea
\ee
For the various orbifold theories based on the torus as the double cover, we
simply need to specialize the general equations obtained in the previous
subsection.

\bigskip

\noindent
$\bullet\,\underline{\rm Klein~Bottle}$

The Klein bottle $K^2$ can be obtained from the torus with modular parameter
$\tau=2i\tau_2$ under the orbifold generated by the parity operator $\Omega'$
introduced at the beginning of section~\ref{sec.orbi}. This operation preserves
the diagonal $U(1)$ current of the conformal field theory which is the
generator of translations along the target space circle. There are two types of
orbifolds that one can form. For the $\Omega'CT$ quotient we obtain an
untwisted theory with partition function
\be
Z_{K^2}(\tau_2)=
\frac{1}{2(2\tau_2)^{1/4}\,\Bigl|\eta(2i\tau_2)\Bigr|}\,
\sum_{m=0}^{pq/2-1}~\sum_{s=-\infty}^\infty
\exp\left\{-\frac{4\pi\tau_2}k\,\Bigl(ps+m\Bigr)^2\right\} \ ,
\lb{ZKu}
\ee
while for $\Omega'T$ we obtain a twisted theory with partition function
\be
Z_{\tilde{K}^2}(\tau_2)=
\frac{\sqrt{k}\,(2\tau_2)^{1/4}}{2\Bigl|\eta(2i\tau_2)\Bigr|}~
\sum_{n=0}^{pq/2-1}~\sum_{s=-\infty}^\infty\exp\left\{-\frac{4\pi\tau_2}
k\,\left(ps+\frac{k}{4}\,n\right)^2\right\} \ .
\lb{ZKt}
\ee

The partition functions (\ref{ZKu}) and (\ref{ZKt}) are related by a modular
transformation of the covering torus $T^2$. To see this, we replace the modulus
$\tau$ in (\ref{ZKt}) by the new modulus
\be
\tilde\tau=-\frac1{2\tau} \ ,
\label{KKmodrel}
\ee
and use the Poisson resummation formula
\be
\sum_{r=-\infty}^\infty\e^{-\pi gr^2-2\pi ibr}=\frac1{\sqrt g}\,
\sum_{s=-\infty}^\infty\e^{-\pi(s-b)^2/g} \ .
\label{Poisson}
\ee
Applying (\ref{Poisson}) twice (in $n$ and $s$) to the partition function
(\ref{ZKt}), and using the modular transformation property
\be
\eta\left(-\frac1\tau\right)=\sqrt{-i\tau}~\eta(\tau)
\label{etamod}
\ee
of the Dedekind function (\ref{Dedekind}), we arrive at the relationship
\be
Z_{\tilde{K}^2}(\tau)=Z_{K^2}(\tilde{\tau}) \ .
\lb{ZKuZKt}
\ee
The surface $\tilde K^2$ exhibits the Klein bottle $K^2$ as a tube terminating
at two cross-caps.

\bigskip

\noindent
$\bullet\,\underline{\rm Annulus}$

We now start considering open orbifold surfaces. The annulus $C^2$ is generated
by the orbifold of the torus $T^2$ with modular parameter $\tau=i\tau_2$ under
the usual worldsheet parity operator $\Omega$. Again, there are two types of
orbifolds. For the $\Omega CT$ quotient we obtain a theory with Neumann
boundary conditions and partition function
\be
Z_{C^2}(\tau_2)=
\frac{1}{2\tau_2^{1/4}\,\Bigl|\eta(i\tau_2)\Bigr|}\,\sum_{m=0}^{pq/2-1}~
\sum_{s=-\infty}^\infty\exp\left\{-\frac{2\pi\tau_2}k\,
\Bigl(ps+m\Bigr)^2\right\} \ ,
\lb{ZCn}
\ee
while for $\Omega T$ we get a theory with Dirichlet boundary conditions and
partition function
\be
Z_{\tilde{C}^2}(\tau_2)=
\frac{\sqrt{k}\,\tau_2^{1/4}}{2\Bigl|\eta(i\tau_2)\Bigr|}~
\sum_{n=0}^{pq/2-1}~\sum_{s=-\infty}^\infty\exp\left\{-\frac{2\pi\tau_2}
k\left(ps+\frac{k}{4}\,n\right)^2\right\} \ .
\lb{ZCd}
\ee
Once again the two partition functions are related by a modular transformation.
Replacing the modulus dependence in (\ref{ZCd}) with
\be
\tilde{\tau}=-\frac{2}{\tau} \ ,
\ee
and using the modular transform (\ref{etamod}), after two Poisson resummations
(in $n$ and $s$) we arrive at the relation
\be
Z_{\tilde{C}^2}(\tau)=Z_{C^2}(\tilde{\tau}) \ .
\ee
The surface $\tilde C^2$ is a cylinder with two boundaries.

\vfill\newpage

\noindent
$\bullet\,\underline{\textrm{M\"obius~Strip}}$

Finally, the M\"obius strip $M^2$ is obtained from the torus with
$\tau=1/2+i\tau_2$ under the orbifold by the parity operation $\Omega$. For the
$\Omega CT$ orbifold one has Neumann boundary conditions and partition function
\be
Z_{M^2}(\tau_2)=\frac{1}{2\tau_2^{1/4}\,\Bigl|\eta(\frac12+i\tau_2)\Bigr|}\,
\sum_{m=0}^{pq/2-1}~\sum_{s=-\infty}^\infty\exp\left\{\frac{\pi i}k\,
\Bigl(1+2i\tau_2\Bigr)\Bigl(ps+m\Bigr)^2\right\} \ ,
\lb{ZMn}
\ee
while for $\Omega T$ we obtain Dirichlet boundary conditions and partition
function
\be
Z_{\tilde{M}^2}(\tau_2)=
\frac{\sqrt{k}\,\tau_2^{1/4}}{2\Bigl|\eta(\frac12+i\tau_2)\Bigr|}\,
\sum_{n=0}^{pq/2-1}~\sum_{s=-\infty}^\infty\exp\left\{\frac{\pi i}k\,
\Bigl(1+2i\tau_2\Bigr)\left(ps+\frac{k}{4}\,n\right)^2\right\} \ .
\lb{ZMd}
\ee
In this case the modular transformation (\ref{KKmodrel}) yields the relation
\be
Z_{\tilde{M}^2}(\tau)=Z_{M^2}(\tilde{\tau}) \ .
\lb{ZMnZMd}
\ee
The surface $\tilde M^2$ is the usual one-loop double twist diagram of open
string theory.

\subsection{\label{subsec.modinv}Remarks on Modular Invariance}

In this subsection we briefly comment on the structure of modular invariant
partition functions constructed from the results obtained above. A more
precise examination of modular invariant quantities will be presented
elsewhere~\cite{progress}. The basic point we wish to make here is that
while the torus partition function (\ref{T2part}) is a modular invariant of
$T^2$, the modular transformations above of the orbifold partition functions
map the direct-channel picture, corresponding to $\Omega CT$ type orbifolds
with partition functions~(\r{ZKu}),~(\r{ZCn}) and~(\r{ZMn}), to the
transverse-channel picture corresponding to $\Omega T$ type orbifolds with
partition functions given by~(\r{ZKt}),~(\r{ZCd}) and~(\r{ZMd}). To build a
modular invariant partition function $Z_{\rm inv}$, it is necessary to sum over
both twisted and untwisted sectors or both sectors corresponding to Neumann and
Dirichlet boundary conditions to get~\cite{POL_1}
\be
Z_{\rm inv}=Z_{\Sigma^{\Omega C}}+Z_{\Sigma^\Omega} \ .
\label{Zinv}
\ee
Note that here, in order to avoid half-integer state multiplicities,
we must {\it add} the statistical sums rather than averaging them. In
the case of open strings, the partition function (\ref{Zinv}) should of
course be understood only as a formal mathematical object which is invariant
under modular transformations, because the theories corresponding to Neumann
and Dirichlet boundary conditions live in different Hilbert spaces. Put
differently, the direct channel amplitudes have character expansions with
integer-valued coefficients which count the number of allowed open and
unoriented string states, and thereby have natural interpretations as
partition functions. On the other hand, in the transverse channel they
correspond to amplitudes of a closed string reflecting from the two
boundaries or crosscaps at the ends of the cylinder and in general
contain non-integer character expansion coefficients. Thus the transverse
channel amplitudes do not count states and have no interpretation as
partition functions. It is not natural to require modular invariance
of these amplitudes, as they have very different physical interpretations.

Nevertheless, the function (\ref{Zinv}) could
have a more profound physical significance if one could find some sort of
tunneling mechanism which interpolates between the two types of boundary
conditions. This would have to be a mechanism which allowed a flow between
superselection sectors of the Hilbert space. A physical system which
possesses the same sort of characteristics is a proton-antiproton system.
While the proton and antiproton correspond to different superselection sectors,
a composite wavefunction corresponding to a proton-antiproton bound state
may be formed. Of course the present open string case is somewhat different
in nature, because here we don't have the analog of charge conjugation as
we tunnel between the different sectors. We are not aware at present of any
such process in open string theory.

The partition function (\ref{Zinv}) corresponds to a modification of the
orthogonal projection in (\ref{piOmega1}) to an operator which picks out
modular invariant states. The appropriate operator acting on states of the
three-dimensional gauge theory is given by
\be
{\cal P}_{\rm inv}=\frac12\,\Bigl(\id+\Omega\,C\,T\Bigr)+
\frac12\,\Bigl(\id+\Omega\,T\Bigr) \ .
\label{piinvOmega}\ee
Note that (\ref{piinvOmega}) is {\it not} a projector. It can only be
completed into an orthogonal projection by adding to (\ref{Zinv}) a
contribution corresponding to the twisted torus states. In general, one
cannot build a modular invariant partition function using projection
operators, but rather only one which is invariant under the relative modular
group of the Riemann surface~\cite{relmod}. This latter group is, however,
trivial at one-loop order. The operator (\ref{piinvOmega}) also ensures that
the partition function is explicitly invariant under
T-duality, which at the level of the worldsheet theory exchanges winding and
Kaluza-Klein modes, and therefore necessarily interchanges both pictures and
partition functions. Furthermore, in addition to the prefactors computed above,
in the case of open string worldsheets there should appear an extra overall
multiplicative factor $N^2$, where $N$ is the number of Chan-Paton factors in
the corresponding string theoretic photon Wilson line~\cite{BPSSym}. This issue
is under current development~\cite{progress}.

\vspace{5mm}\noindent {\large\bf Acknowledgments}

We thank M.~Bianchi and A.~Sagnotti for helpful comments on the manuscript.
P.C.F. is grateful for the hospitality of the Department of Mathematics of
Heriot-Watt University, where this work was completed. P.C.F. and I.I.K. thank
A.~Kovner for helpful discussions. The work of P.C.F. is supported by Grant
SFRH/BPD/5638/2001 from FCT~(Portugal). The work of I.I.K. is supported by
PPARC Grant PPA/G/0/1998/00567 and EU Grant FMRXCT960090. The work of R.J.S.
is supported by a PPARC Advanced Fellowship.

\end{document}